\documentclass[10pt]{article}
\usepackage{amsmath}
\usepackage{amssymb}
\usepackage{graphicx}
\usepackage{color}
\usepackage[table,xcdraw]{xcolor}
\usepackage{etoolbox}
\usepackage{orcidlink}
\usepackage{hyperref}
\hypersetup{colorlinks=true, linkcolor=blue, citecolor=blue, urlcolor=blue}
\usepackage{array}
\usepackage{longtable}
\usepackage{booktabs}
\usepackage{setspace}
\usepackage{colortbl}
\usepackage{subcaption}
\usepackage{geometry}
\usepackage{colortbl,xcolor}
\usepackage{tikz}
\usepackage{array}
\usepackage{diagbox}
\RequirePackage[numbers,sort&compress]{natbib}
\begin{document}
\baselineskip=20pt
\begin{center}
\setstretch{1.8}
{\LARGE {\bf Aharonov-Bohm Effect for Cooper Pairs in Kerr Spacetime: Gravitomagnetic Phase Shifts from Frame Dragging}}
\end{center}
\vspace{0.1cm}
\begin{center}
{\bf Erdem Sucu}\orcidlink{0009-0000-3619-1492}\\
Physics Department, Eastern Mediterranean University, Famagusta 99628, North Cyprus via Mersin 10, Turkey\\
e-mail: erdemsc07@gmail.com\\[0.3cm]
{\bf \.{I}zzet Sakall{\i}}\orcidlink{0000-0001-7827-9476}\\
Physics Department, Eastern Mediterranean University, Famagusta 99628, North Cyprus via Mersin 10, Turkey\\
e-mail: izzet.sakalli@emu.edu.tr (Corresponding author)
\end{center}
\vspace{0.2cm}
\begin{abstract}
\setstretch{1.4}
The unification of quantum mechanics and general relativity remains among the most profound challenges in fundamental physics. Here we investigate a novel quantum probe of strong-field gravity: the gravitomagnetic Aharonov-Bohm (AB) effect for Cooper pairs propagating in Kerr spacetime. The frame-dragging induced by a rotating black hole (BH) generates an effective vector potential through the off-diagonal metric component $g_{t\phi}$, which couples to the macroscopic phase of the superconducting condensate. We derive the gauge-invariant AB phase shift $\Delta\theta = (4\pi m^* Ma/\hbar)(1/r_2 - 1/r_1)$ for an interferometer with arms at radii $r_1$ and $r_2$, where $m^* = 2m_e$ is the Cooper pair mass and $a$ is the BH spin parameter. Remarkably, the predicted phases reach $|\Delta\theta| \sim 10^{24}$ radians for Sgr~A* and $\sim 10^{27}$ radians for M87*, reflecting the enormous gravitomagnetic flux near supermassive BHs. We analyze the dependence on interferometer geometry, demonstrate that tidal disruption of Cooper pairs is negligible at distances $r \gtrsim 10\,r_s$, and establish connections to the geometric Berry phase. Although direct experimental realization remains beyond current technology due to the vast distances involved, our framework provides quantitative predictions linking quantum coherence to spacetime curvature, complementing recent observations of gravitational AB phases in atom interferometry.
\end{abstract}
\pagebreak
\tableofcontents
\pagebreak

{\color{black}

\section{Introduction} \label{sec1}

The interface between quantum mechanics and general relativity remains one of the most profound frontiers in fundamental physics. While both theories have been spectacularly confirmed within their respective domains---quantum mechanics governing the microscopic world of atoms and particles, general relativity describing the large-scale structure of spacetime---their unification continues to elude a complete theoretical framework \cite{Kiefer:2007}. Experimental probes that simultaneously involve both quantum coherence and strong gravitational effects are therefore of paramount importance, as they may reveal new physics or provide stringent tests of our current understanding.

The AB effect stands as one of the most striking demonstrations of the physical significance of gauge potentials in quantum mechanics \cite{Aharonov:1959}. First predicted by Ehrenberg and Siday \cite{Ehrenberg:1949} and independently analyzed by Aharonov and Bohm, the effect shows that charged particles acquire a measurable phase shift when encircling a region of magnetic flux, even when traveling through field-free regions where the Lorentz force vanishes. This topological phase, proportional to the enclosed flux, was confirmed experimentally by Chambers \cite{Chambers:1960} and definitively demonstrated using electron holography by Tonomura and collaborators \cite{Tonomura:1986}. The AB effect fundamentally changed our understanding of electromagnetism, establishing that potentials---not just fields---have direct physical consequences in quantum theory \cite{Peshkin:1989}.

The gravitational analogue of the AB effect has attracted theoretical attention since the pioneering work of DeWitt \cite{DeWitt:1966}, who showed that superconductors could in principle respond to gravitational fields through their macroscopic quantum coherence. In the weak-field limit, the gravitomagnetic field generated by a rotating mass acts analogously to a magnetic field for matter waves, with the metric component $g_{t\phi}$ playing the role of a vector potential \cite{Mashhoon:1993,Ruggiero:2002,Mashhoon:2003}. This gravitoelectromagnetic analogy has proven remarkably fruitful, providing intuitive understanding of frame-dragging effects and guiding experimental searches for gravitomagnetic phenomena \cite{Ciufolini:2004,Everitt:2011}.

A major experimental breakthrough occurred in 2022 when Overstreet \textit{et al.}\ reported the first observation of a gravitational AB effect using atom interferometry \cite{Overstreet:2022}. By splitting ultracold $^{87}$Rb atoms into spatially separated wave packets and subjecting one arm to the gravitational potential of a kilogram-scale tungsten mass, they detected phase shifts consistent with the gravitational AB prediction. This landmark result, achieved with phase sensitivities of $\sim 10^{-4}$ rad, demonstrated that gravity induces AB-type phase shifts analogous to electromagnetic interactions and opened new avenues for precision measurements of Newton's constant $G$ \cite{Roura:2022,Georgescu:2023}. Subsequent theoretical work has explored extensions to microgravity environments \cite{Liu:2024} and the scalar gravitational AB effect using atomic clocks in eccentric orbits \cite{Tobar:2024}.

Parallel to these developments in matter-wave interferometry, the Event Horizon Telescope (EHT) has achieved direct imaging of supermassive BHs (SMBHs), providing unprecedented access to the strong-field regime of general relativity \cite{EHT:2019,EHT:2022}. The images of M87* and Sgr~A* reveal the characteristic ``shadow'' predicted by Kerr geometry and have enabled initial constraints on BH masses and spins \cite{EHT:2019b,Psaltis:2020,Vagnozzi:2023}. Recent analyses suggest that M87* spins at approximately $80\%$ of the theoretical maximum ($a/M \gtrsim 0.8$) \cite{Cui:2023,Drew:2025}, while observations of jet precession with an $\sim 11$-year period provide independent evidence for a rapidly rotating Kerr BH with a tilted accretion disk \cite{Cui:2023Nature}. For Sgr~A*, spin estimates remain more uncertain but favor moderate to high values \cite{GravityCollaboration:2020,Dokuchaev:2023}. These observations confirm that astrophysical BHs possess substantial angular momenta and thus generate significant gravitomagnetic fields---precisely the conditions required for a strong gravitational AB effect.

The gravitomagnetic field of a Kerr BH is encoded in the off-diagonal metric component $g_{t\phi}$, which describes the ``dragging of inertial frames'' first predicted by Lense and Thirring \cite{LenseThirring:1918}. This frame-dragging effect has been measured in the weak-field regime by Gravity Probe B \cite{Everitt:2011} and the LAGEOS/LARES satellite laser ranging experiments \cite{Ciufolini:2004,Ciufolini:2019,Ciufolini:2023}, confirming general relativistic predictions at the level of a few percent. Near a Kerr BH, however, the gravitomagnetic potential is enormously amplified: at the innermost stable circular orbit of a maximally spinning BH, the frame-dragging angular velocity approaches $\Omega \sim c/(2r_+)$, where $r_+$ is the horizon radius. This strong-field enhancement makes rotating BHs ideal laboratories for studying gravitomagnetic effects at their most extreme.

Cooper pairs in superconductors offer a unique probe of gravitomagnetic fields due to their macroscopic quantum coherence. Unlike single-particle matter waves, Cooper pairs form a Bose-Einstein condensate described by the Ginzburg-Landau (GL) order parameter $\Psi = |\Psi|e^{i\theta}$, whose phase $\theta$ extends coherently over macroscopic distances \cite{Tinkham:2004,Poole:2007}. This coherence underlies the remarkable sensitivity of superconducting quantum interference devices (SQUIDs), which can detect magnetic flux changes as small as $10^{-6}\,\Phi_0$ per $\sqrt{\text{Hz}}$, where $\Phi_0 = h/(2e) = 2.07 \times 10^{-15}$ Wb is the magnetic flux quantum \cite{Clarke:2004}. The theoretical framework for coupling superconductors to gravitational fields was developed by DeWitt \cite{DeWitt:1966}, Papini \cite{Papini:1967}, and Anandan \cite{Anandan:1977}, establishing that the Cooper pair phase acquires contributions from both electromagnetic and gravitational potentials.

The link between superconductivity and gravity has been explored in various contexts, including rotating superconductors exhibiting the London moment \cite{London:1950,Tate:1989}, proposed gravitational wave detectors using superconducting circuits \cite{Inan:2017}, and theoretical studies of superconductor-gravity coupling in linearized gravity \cite{Modanese:1996,Ummarino:2022}. While some experimental claims of anomalous gravitational effects in superconductors \cite{Podkletnov:1992} remain controversial and unconfirmed, the underlying theoretical framework connecting GL theory to curved spacetime is well established \cite{Anandan:1994,Zhang:2026lrd}. Recent work has also explored quantum field theory in curved spacetime using ultracold atomic systems as analog gravity simulators \cite{Tajik:2023,Barcelo:2011}, demonstrating that laboratory systems can probe aspects of quantum mechanics in curved backgrounds. In the context of holography, Hartman and Hartnoll \cite{Hartman:2010} showed that a four-fermion contact interaction between charged Dirac fermions can induce Cooper pairing and a superconducting instability near a charged Anti-de Sitter black hole. Their analysis yielded an analytic formula for the critical temperature, $T_c \propto \mu \exp(-M_F^2 L^2/N_{\rm eff})$, where $\mu$ is the chemical potential, $M_F$ the interaction energy scale, $L$ the AdS radius, and $N_{\rm eff}$ the effective density of states at the Fermi surface. The instability occurs only when the fermion dispersion is linear near the Fermi surface, with $T_c \to 0$ as the marginal Fermi liquid regime is approached. While their work employed holographic methods in AdS spacetimes with electrically charged black holes, the underlying physics of Cooper pair formation in curved backgrounds provides valuable theoretical context for our analysis of the gravitomagnetic AB effect in asymptotically flat Kerr spacetime.

In this paper, we develop the theoretical framework for the gravitomagnetic AB effect experienced by Cooper pairs in Kerr spacetime. Our primary contributions are: (i) derivation of the gauge-invariant AB phase for a Cooper pair interferometer orbiting a Kerr BH, expressed in terms of the mass $M$ and spin parameter $a$; (ii) analysis of how the phase depends on interferometer geometry, including equatorial versus inclined configurations; (iii) numerical evaluation of the phase for astrophysical BHs including Sgr~A*, M87*, and stellar-mass systems; (iv) assessment of experimental challenges including tidal forces, thermal environment, and detection sensitivity; and (v) discussion of the connection to geometric (Berry) phases and potential analog gravity realizations.

The motivation for this study is threefold. First, the Cooper pair AB phase provides a quantum-coherent observable that directly encodes BH spin, complementing electromagnetic and gravitational-wave measurements. While the enormous distances to astrophysical BHs preclude near-term experimental realization, the theoretical framework establishes quantitative predictions against which future technologies can be assessed. Second, the analysis illuminates the deep structural parallels between electromagnetic and gravitomagnetic gauge theories at the quantum level, with the metric component $g_{t\phi}$ playing the precise role of the electromagnetic vector potential $A_\phi$. Third, the study contributes to the broader program of understanding quantum coherence in curved spacetime, a topic of growing interest given recent experimental advances in atom interferometry \cite{Overstreet:2022}, optical clocks \cite{Bothwell:2022}, and proposals for testing the quantum nature of gravity \cite{Bose:2017,Marletto:2017}.

The paper is organized as follows. In Sec.~\ref{sec2}, we review the Kerr metric in Boyer-Lindquist coordinates and extract the gravitomagnetic vector potential, presenting numerical values and visualizations of the relevant metric functions. Section~\ref{sec3} develops the GL formalism in curved spacetime and derives the gauge-covariant phase evolution, including a comparison with the electromagnetic AB effect and analysis of coherence requirements. In Sec.~\ref{sec4}, we compute the AB phase for various interferometer geometries, examining the dependence on arm separation, inclination angle, and BH parameters. Section~\ref{sec5} discusses observational prospects, comparing with existing gravitomagnetic measurements, analyzing tidal and thermal constraints, and evaluating laboratory analog systems. We conclude in Sec.~\ref{sec6} with a summary and outlook for future investigations.

Throughout this paper, we use geometrized units with $G = c = 1$ unless otherwise stated, and metric signature $(-,+,+,+)$. Physical units are restored when presenting numerical results. The Cooper pair mass is $m^* = 2m_e = 1.82 \times 10^{-30}$ kg, and the Cooper pair charge is $e^* = 2e = 3.20 \times 10^{-19}$ C.

\section{Kerr Geometry and Gravitomagnetic Potential} \label{sec2}

In this section, we establish the geometric framework for analyzing the gravitomagnetic AB effect in Kerr spacetime. The rotating BH solution discovered by Kerr \cite{Kerr:1963} contains an off-diagonal metric component $g_{t\phi}$ that encodes the frame-dragging phenomenon. This component plays a role analogous to the electromagnetic vector potential in the standard AB effect, coupling to matter waves and inducing observable phase shifts. We begin by reviewing the Kerr metric in BL coordinates, then extract the gravitomagnetic potential from the ADM decomposition. The gravitomagnetic flux through closed loops is computed, and the frame-dragging angular velocity is analyzed. These geometric quantities will serve as the foundation for the Cooper pair phase calculations developed in subsequent sections.

\subsection{Boyer-Lindquist Coordinates}

The Kerr metric describing a rotating BH of mass $M$ and angular momentum $J = Ma$ takes the form in BL coordinates $(t, r, \vartheta, \phi)$ \cite{Kerr:1963,Boyer:1967}:
\begin{equation}\label{eq:kerr_metric}
ds^2 = -\left(1 - \frac{2Mr}{\Sigma}\right)dt^2 - \frac{4Mar\sin^2\vartheta}{\Sigma}\,dt\,d\phi + \frac{\Sigma}{\Delta}\,dr^2 + \Sigma\,d\vartheta^2 + \frac{\mathcal{A}\sin^2\vartheta}{\Sigma}\,d\phi^2,
\end{equation}
where the metric functions read
\begin{align}
\Sigma &= r^2 + a^2\cos^2\vartheta, \label{eq:Sigma}\\
\Delta &= r^2 - 2Mr + a^2, \label{eq:Delta}\\
\mathcal{A} &= (r^2 + a^2)^2 - a^2\Delta\sin^2\vartheta. \label{eq:A_func}
\end{align}
The event horizons are located at $r_{\pm} = M \pm \sqrt{M^2 - a^2}$, and the existence of horizons requires $|a| \leq M$. The outer horizon $r_+$ coincides with the inner horizon $r_-$ at $r = M$ when the BH reaches the extremal limit $a = M$.

At the equatorial plane ($\vartheta = \pi/2$), the metric functions simplify considerably. We have $\Sigma = r^2$, and the relevant metric components become
\begin{equation}
g_{tt}\big|_{\rm eq} = \frac{2M - r}{r}, \qquad g_{t\phi}\big|_{\rm eq} = -\frac{2Ma}{r}, \qquad g_{\phi\phi}\big|_{\rm eq} = \frac{(2M + r)a^2 + r^3}{r}.
\end{equation}
This simplified form will be central to our analysis of the AB phase for equatorial interferometers.

\subsection{Gravitomagnetic Vector Potential}

The ADM decomposition of the metric provides a natural framework for identifying the gravitomagnetic potential \cite{Arnowitt:1962}. The line element can be written as
\begin{equation}\label{eq:ADM}
ds^2 = -N^2 dt^2 + \gamma_{ij}(dx^i + N^i dt)(dx^j + N^j dt),
\end{equation}
where $N$ denotes the lapse function and $N^i$ is the shift vector. Comparing with Eq.~\eqref{eq:kerr_metric}, we find that the shift vector has a single nonzero component given by
\begin{equation}\label{eq:shift}
N^\phi = -\frac{g_{t\phi}}{g_{\phi\phi}} = \frac{2Mar}{\mathcal{A}}.
\end{equation}
At the equatorial plane, this reduces to
\begin{equation}
N^\phi\big|_{\rm eq} = \frac{2Ma}{(2M + r)a^2 + r^3}.
\end{equation}

This shift vector represents the angular velocity at which local inertial frames are dragged by the rotating BH. An observer who remains at fixed spatial coordinates in BL coordinates must rotate with angular velocity $\Omega_{\text{ZAMO}} = N^\phi$ relative to distant stars. These Zero Angular Momentum Observers (ZAMOs) play an important role in the physical interpretation of frame dragging \cite{Visser:2008}.

In the weak-field, slow-rotation regime where $r \gg M$ and $a \ll r$, the gravitomagnetic effects can be described by an analogy with electromagnetism \cite{Mashhoon:2003}. The gravitomagnetic vector potential takes the form $\mathbf{A}_g = \frac{1}{2}\mathbf{g} \times \mathbf{r}$, where $\mathbf{g}$ is the gravitomagnetic field. For the full Kerr geometry beyond this approximation, we identify the effective gravitomagnetic potential through the metric component
\begin{equation}\label{eq:Ag_Kerr}
A_g^\phi \equiv g_{t\phi} = -\frac{2Mar\sin^2\vartheta}{\Sigma}.
\end{equation}

This identification allows us to treat the frame-dragging contribution as a gauge potential for matter waves, directly analogous to the electromagnetic vector potential in the standard AB effect. The key difference is that the gravitomagnetic potential couples to mass rather than charge.

Figure~\ref{fig:gtphi_r} displays the radial dependence of $g_{t\phi}$ at the equatorial plane for various spin parameters ranging from $a/M = 0.1$ to $a/M = 0.99$. The gravitomagnetic potential is negative throughout the exterior region, reflecting the dragging of inertial frames in the direction of BH rotation. As expected, $|g_{t\phi}|$ increases with the spin parameter $a$ and decreases with radial distance, following the $1/r$ falloff at large distances.

\begin{figure}[htbp]
\centering
\includegraphics[width=0.75\textwidth]{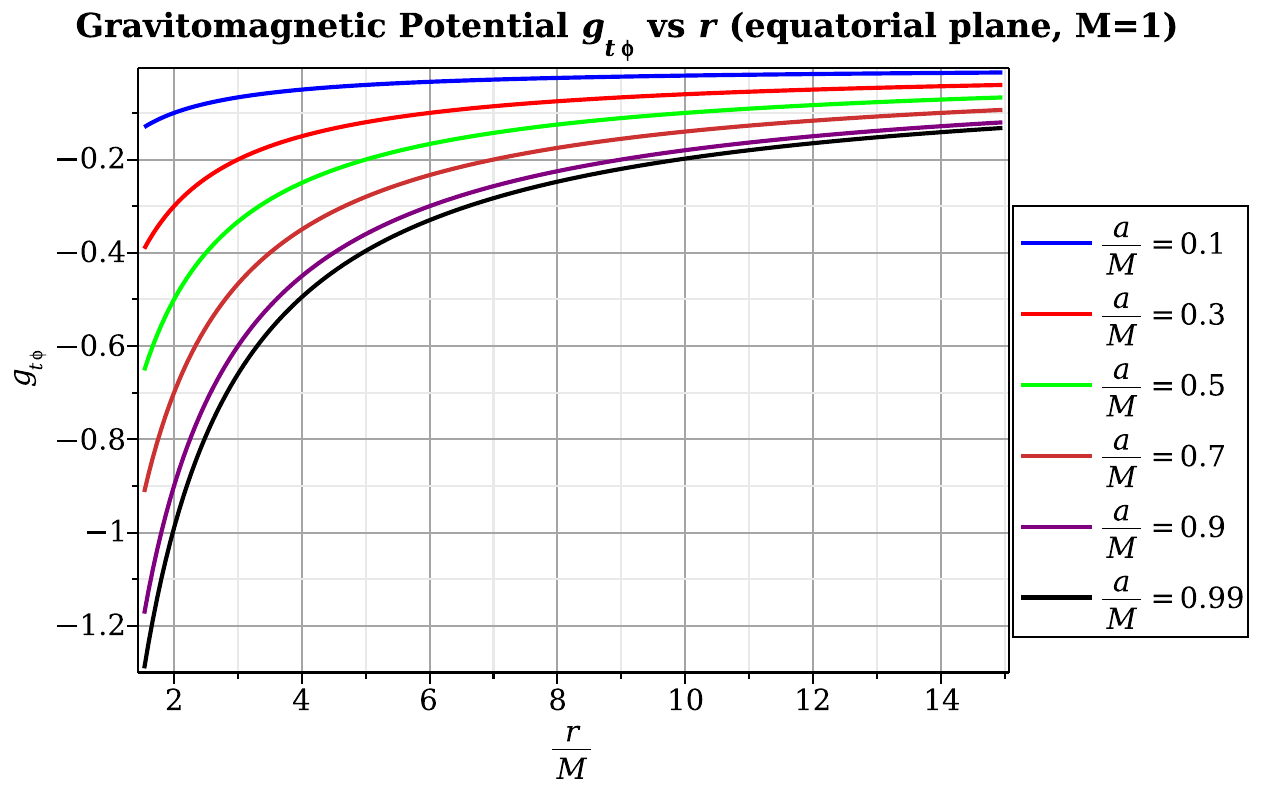}
\caption{Gravitomagnetic potential $g_{t\phi}$ as a function of $r/M$ at the equatorial plane for spin parameters $a/M = 0.1, 0.3, 0.5, 0.7, 0.9$, and $0.99$. The potential becomes more negative (stronger frame dragging) for higher spin and smaller radii. At large $r$, all curves approach zero as $g_{t\phi} \sim -2Ma/r$.}
\label{fig:gtphi_r}
\end{figure}

The angular dependence of $g_{t\phi}$ is shown in Fig.~\ref{fig:gtphi_theta} for fixed spin parameter $a/M = 0.9$ and various radii. The $\sin^2\vartheta$ factor in Eq.~\eqref{eq:Ag_Kerr} produces a characteristic profile that vanishes at the poles ($\vartheta = 0, \pi$) and reaches maximum magnitude at the equatorial plane ($\vartheta = \pi/2$). This angular structure has direct consequences for interferometer design: equatorial configurations maximize the gravitomagnetic signal, while polar orbits experience no net AB phase.

\begin{figure}[htbp]
\centering
\includegraphics[width=0.75\textwidth]{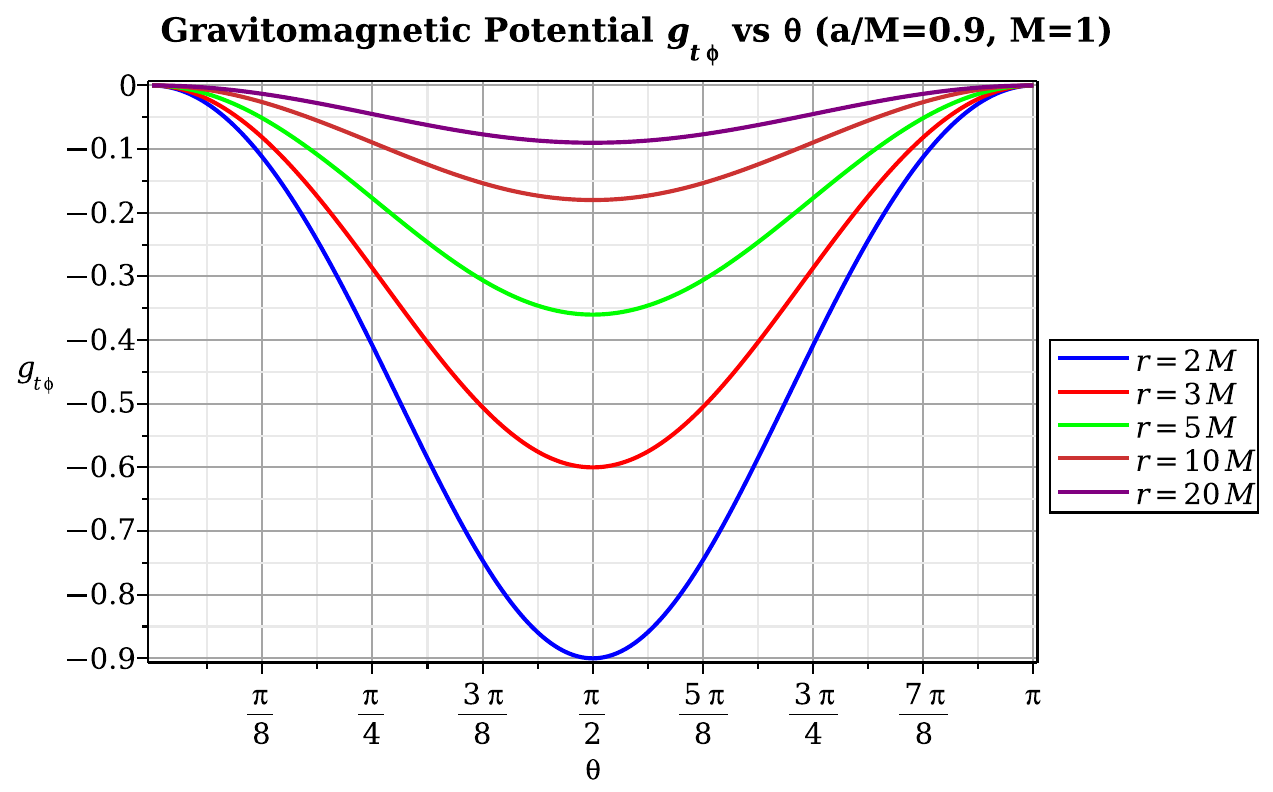}
\caption{Angular dependence of the gravitomagnetic potential $g_{t\phi}$ for $a/M = 0.9$ at radii $r = 2M, 3M, 5M, 10M$, and $20M$. The potential vanishes at the poles and reaches its extremum at the equatorial plane, following the $\sin^2\vartheta$ dependence.}
\label{fig:gtphi_theta}
\end{figure}

Table~\ref{tab:gtphi_values} presents numerical values of $g_{t\phi}$ at the equatorial plane for various combinations of radius and spin parameter. These values confirm the linear scaling with $a$ and inverse scaling with $r$ evident from the analytical expression $g_{t\phi}|_{\rm eq} = -2Ma/r$.

\begin{table}[ht!]
\centering
\caption{Numerical values of the gravitomagnetic potential $g_{t\phi}$ at the equatorial plane for various radii and spin parameters. We set $M=1$ in geometric units.}
\label{tab:gtphi_values}
\begin{tabular}{c|cccc}
\hline\hline
\rowcolor{orange!50}
$r/M$ & $a/M=0.1$ & $a/M=0.5$ & $a/M=0.9$ & $a/M=0.99$ \\
\hline
$2$ & $-0.10000$ & $-0.50000$ & $-0.90000$ & $-0.99000$ \\
$3$ & $-0.06667$ & $-0.33334$ & $-0.60000$ & $-0.66000$ \\
$4$ & $-0.05000$ & $-0.25000$ & $-0.45000$ & $-0.49500$ \\
$5$ & $-0.04000$ & $-0.20000$ & $-0.36000$ & $-0.39600$ \\
$6$ & $-0.03333$ & $-0.16667$ & $-0.30000$ & $-0.33000$ \\
$8$ & $-0.02500$ & $-0.12500$ & $-0.22500$ & $-0.24750$ \\
$10$ & $-0.02000$ & $-0.10000$ & $-0.18000$ & $-0.19800$ \\
$15$ & $-0.01333$ & $-0.06667$ & $-0.12000$ & $-0.13200$ \\
$20$ & $-0.01000$ & $-0.05000$ & $-0.09000$ & $-0.09900$ \\
\hline\hline
\end{tabular}
\end{table}

Figure~\ref{fig:gtphi_3D} provides a global view of $g_{t\phi}$ in the $(r, \vartheta)$ plane for $a/M = 0.9$. The three-dimensional surface clearly shows how the gravitomagnetic potential concentrates near the equator and close to the horizon, where frame-dragging effects are most pronounced. This visualization aids in identifying optimal regions for positioning gravitomagnetic AB interferometers.

\begin{figure}[htbp]
\centering
\includegraphics[width=0.55\textwidth]{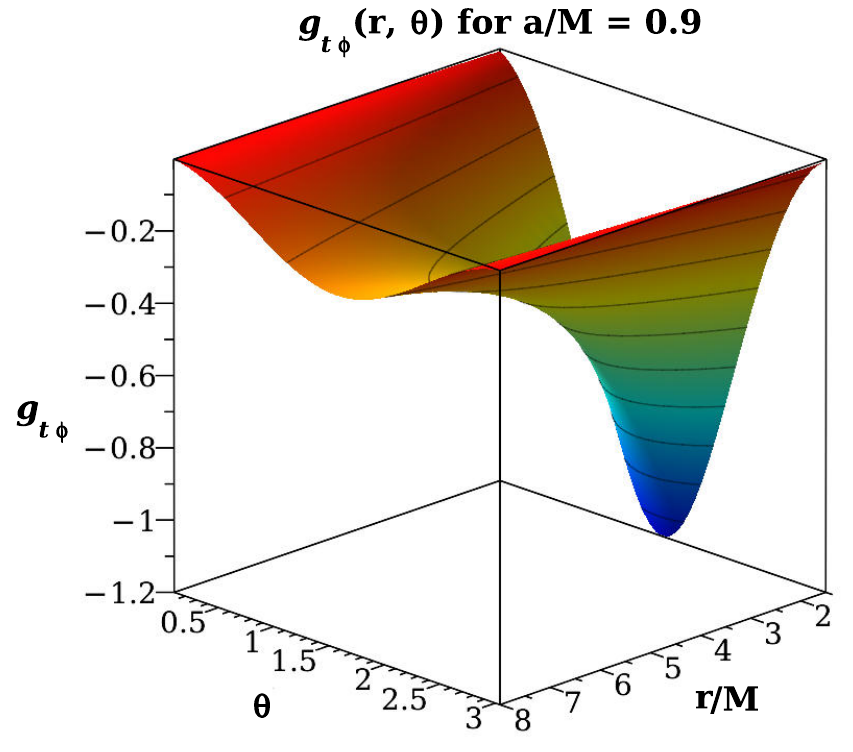}
\caption{Three-dimensional surface plot of the gravitomagnetic potential $g_{t\phi}(r, \vartheta)$ for $a/M = 0.9$. The deep well near $r \approx 2M$ and $\vartheta = \pi/2$ indicates where frame-dragging effects are strongest.}
\label{fig:gtphi_3D}
\end{figure}

\subsection{Gravitomagnetic Flux}

For a closed circular loop $\mathcal{C}$ in the equatorial plane at constant radius $r$, the gravitomagnetic flux is obtained by integrating the potential around the loop:
\begin{equation}\label{eq:flux}
\Phi_g = \oint_{\mathcal{C}} g_{t\phi}\,d\phi = g_{t\phi}\big|_{\vartheta=\pi/2} \cdot 2\pi = -\frac{4\pi Ma}{r}.
\end{equation}
This flux plays the role of the magnetic flux $\Phi_B$ in the electromagnetic AB effect. The gravitomagnetic flux is proportional to the BH angular momentum $J = Ma$ and inversely proportional to the loop radius. Expressing the result in terms of $J$ yields the simple form $\Phi_g = -4\pi J/r$.

The behavior of $\Phi_g$ as a function of radius is shown in Fig.~\ref{fig:flux}. For a rapidly rotating BH with $a/M = 0.99$, the flux at $r = 2M$ reaches $|\Phi_g| \approx 6.22$, while at $r = 10M$ it decreases to $|\Phi_g| \approx 1.24$. This strong radial dependence suggests that interferometer configurations with arms at different radii can generate measurable phase differences.

\begin{figure}[htbp]
\centering
\includegraphics[width=0.75\textwidth]{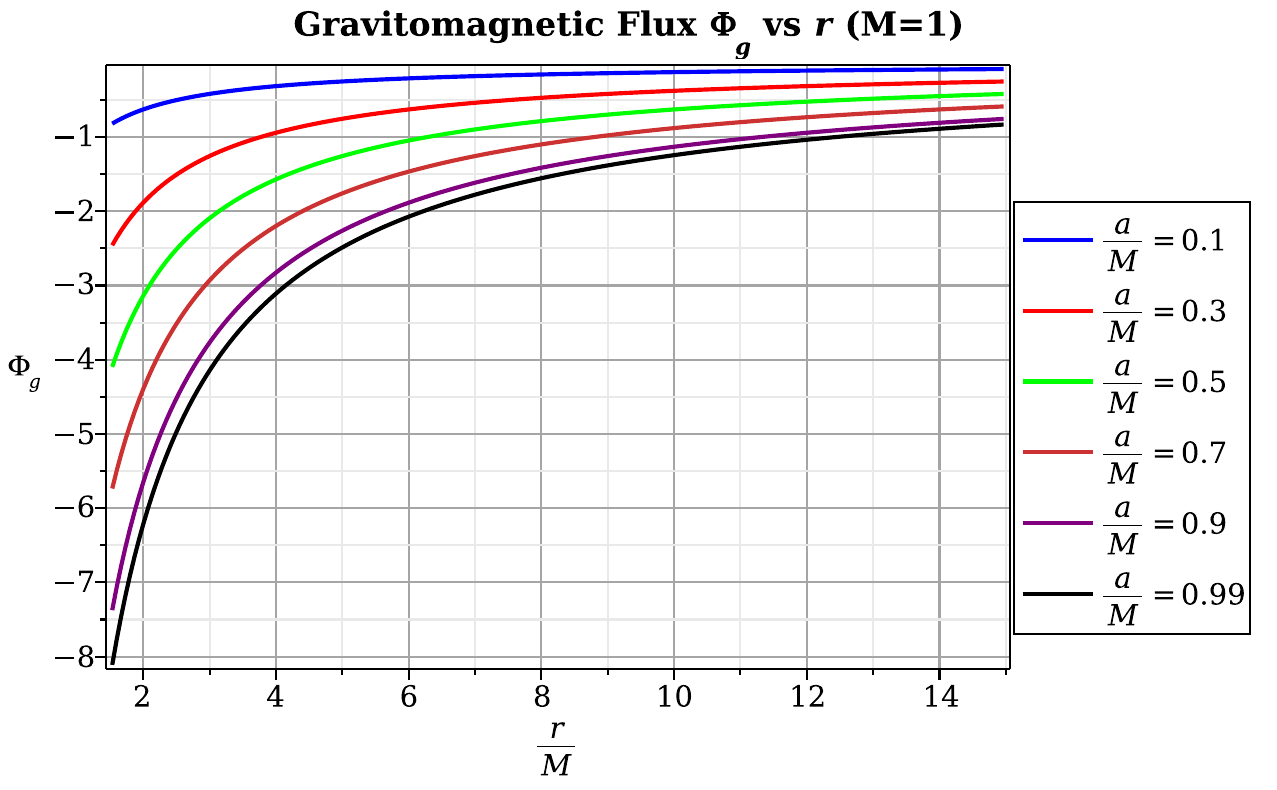}
\caption{Gravitomagnetic flux $\Phi_g$ as a function of $r/M$ for spin parameters $a/M = 0.1, 0.3, 0.5, 0.7, 0.9$, and $0.99$. The flux determines the AB phase shift through the relation $\Delta\theta = (m^*/\hbar)\Phi_g$.}
\label{fig:flux}
\end{figure}

Table~\ref{tab:flux_values} collects numerical values of $\Phi_g$ for the same parameter combinations as Table~\ref{tab:gtphi_values}. These values directly determine the AB phase shift through the relation $\Delta\theta = (m^*/\hbar)\Phi_g$, which we develop in subsequent sections.

\begin{table}[ht!]
\centering
\caption{Gravitomagnetic flux $\Phi_g$ for equatorial circular loops at various radii. The flux is computed via $\Phi_g = 2\pi g_{t\phi}|_{\vartheta=\pi/2}$ with $M=1$.}
\label{tab:flux_values}
\begin{tabular}{c|cccc}
\hline\hline
\rowcolor{orange!50}
$r/M$ & $a/M=0.1$ & $a/M=0.5$ & $a/M=0.9$ & $a/M=0.99$ \\
\hline
$2$ & $-0.62832$ & $-3.14160$ & $-5.65480$ & $-6.22040$ \\
$3$ & $-0.41888$ & $-2.09440$ & $-3.76990$ & $-4.14680$ \\
$4$ & $-0.31416$ & $-1.57080$ & $-2.82740$ & $-3.11020$ \\
$5$ & $-0.25133$ & $-1.25660$ & $-2.26200$ & $-2.48820$ \\
$6$ & $-0.20944$ & $-1.04720$ & $-1.88500$ & $-2.07340$ \\
$8$ & $-0.15708$ & $-0.78540$ & $-1.41370$ & $-1.55510$ \\
$10$ & $-0.12566$ & $-0.62832$ & $-1.13100$ & $-1.24410$ \\
$15$ & $-0.08378$ & $-0.41888$ & $-0.75400$ & $-0.82940$ \\
$20$ & $-0.06283$ & $-0.31416$ & $-0.56548$ & $-0.62204$ \\
\hline\hline
\end{tabular}
\end{table}

\subsection{Frame-Dragging Angular Velocity}

The angular velocity of locally non-rotating observers (ZAMOs) is given by
\begin{equation}\label{eq:Omega_ZAMO}
\Omega_{\text{ZAMO}} = -\frac{g_{t\phi}}{g_{\phi\phi}} = \frac{2Mar}{\mathcal{A}}.
\end{equation}
At the equatorial plane, this reduces to
\begin{equation}\label{eq:Omega_eq}
\Omega_{\text{ZAMO}}\big|_{\vartheta=\pi/2} = \frac{2Ma}{(2M + r)a^2 + r^3}.
\end{equation}

At the outer horizon $r = r_+$, the frame-dragging angular velocity reaches the horizon angular velocity
\begin{equation}\label{eq:Omega_H}
\Omega_H = \frac{a}{2Mr_+} = \frac{a}{2M(M + \sqrt{M^2 - a^2})}.
\end{equation}
For an extremal Kerr BH with $a = M$, this becomes $\Omega_H = 1/(2M)$, representing the maximum possible frame-dragging rate. Figure~\ref{fig:Omega} displays the radial profile of $\Omega_{\text{ZAMO}}$ for various spin parameters. Near the horizon, frame dragging becomes increasingly pronounced as $a/M$ approaches unity.

\begin{figure}[htbp]
\centering
\includegraphics[width=0.75\textwidth]{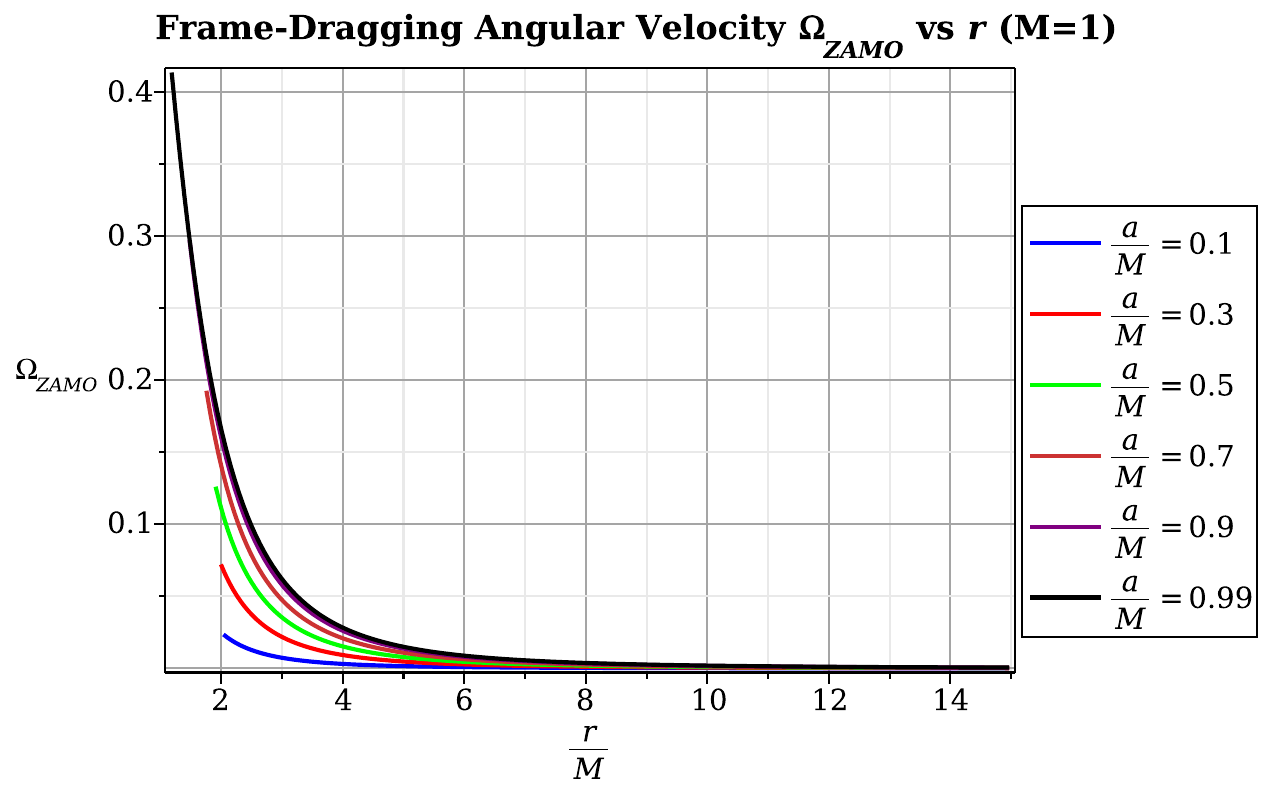}
\caption{Frame-dragging angular velocity $\Omega_{\text{ZAMO}}$ as a function of $r/M$ for spin parameters $a/M = 0.1, 0.3, 0.5, 0.7, 0.9$, and $0.99$. Each curve starts just outside the corresponding horizon radius $r_+ = M + \sqrt{M^2 - a^2}$. For the near-extremal case $a/M = 0.99$, the angular velocity reaches $\Omega \approx 0.42$ near the horizon.}
\label{fig:Omega}
\end{figure}

The connection between the gravitomagnetic potential and the frame-dragging angular velocity can be expressed as $g_{t\phi} = -\Omega_{\text{ZAMO}} g_{\phi\phi}$. This relation shows that the AB phase acquired by a Cooper pair condensate is directly linked to the frame-dragging experienced by local observers, providing a quantum-mechanical probe of this classical relativistic effect.

==============================================================================

\section{Cooper Pair Dynamics in Curved Spacetime} \label{sec3}

Having established the geometric properties of the Kerr spacetime and identified the gravitomagnetic potential, we now turn to the quantum mechanical description of Cooper pairs propagating in this curved background. The superconducting condensate is characterized by a macroscopic wavefunction whose phase is sensitive to both electromagnetic and gravitational influences. In flat spacetime, the Ginzburg-Landau (GL) theory provides the standard framework for describing superconductivity, and its extension to curved spacetime requires careful treatment of the covariant derivative and the coupling between the condensate phase and the spacetime geometry. We derive the phase evolution equation for Cooper pairs in Kerr spacetime and demonstrate that the gravitomagnetic contribution to the phase shift is gauge-invariant, depending only on the enclosed gravitomagnetic flux.

\subsection{Ginzburg-Landau Theory}

The superconducting state is described by a complex order parameter $\Psi(\mathbf{x}) = |\Psi|e^{i\theta}$, where $|\Psi|^2$ represents the density of Cooper pairs and $\theta$ is the macroscopic phase \cite{Ginzburg:1950,Tinkham:2004}. The GL free energy functional governs the behavior of $\Psi$ near the critical temperature, and the resulting GL equation determines the spatial and temporal evolution of the order parameter.

In flat spacetime with electromagnetic coupling, the gauge-covariant derivative acting on $\Psi$ takes the form
\begin{equation}\label{eq:cov_deriv_flat}
D_\mu \Psi = \left(\partial_\mu - \frac{ie^*}{\hbar}A_\mu\right)\Psi,
\end{equation}
where $e^* = 2e = 3.20 \times 10^{-19}$ C is the Cooper pair charge and $A_\mu$ is the electromagnetic four-potential. This minimal coupling ensures gauge invariance of the theory: under the gauge transformation $A_\mu \to A_\mu + \partial_\mu \chi$ and $\Psi \to \Psi e^{ie^*\chi/\hbar}$, the covariant derivative transforms covariantly.

The supercurrent density follows from the GL free energy as
\begin{equation}\label{eq:supercurrent_flat}
\mathbf{J}_s = \frac{e^*\hbar}{2m^* i}\left(\Psi^* \nabla \Psi - \Psi \nabla \Psi^*\right) - \frac{(e^*)^2}{m^*}|\Psi|^2 \mathbf{A},
\end{equation}
where $m^* = 2m_e = 1.82 \times 10^{-30}$ kg is the Cooper pair mass. Writing $\Psi = |\Psi|e^{i\theta}$, this becomes
\begin{equation}\label{eq:supercurrent_phase}
\mathbf{J}_s = \frac{e^* |\Psi|^2}{m^*}\left(\hbar \nabla\theta - e^* \mathbf{A}\right).
\end{equation}
The quantity in parentheses is the gauge-invariant momentum, and the supercurrent vanishes when $\hbar \nabla\theta = e^* \mathbf{A}$.

In curved spacetime, the partial derivative $\partial_\mu$ must be replaced by the covariant derivative $\nabla_\mu$ that accounts for the gravitational connection. For a scalar field like the order parameter magnitude $|\Psi|$, the covariant derivative coincides with the partial derivative. However, the phase gradient and its coupling to the metric require more careful treatment, as we discuss in the following subsection.

\subsection{Phase Evolution in Kerr Spacetime}

The dynamics of a quantum particle in curved spacetime can be analyzed through the WKB approximation, where the wavefunction takes the form $\Psi \propto e^{iS/\hbar}$ with $S$ the classical action \cite{DeWitt:1966,Papini:1967}. For a Cooper pair with mass $m^* = 2m_e$, the phase $\theta = S/\hbar$ evolves according to the Hamilton-Jacobi equation in curved spacetime.

Writing $\Psi = |\Psi|e^{i\theta}$, the phase gradient couples to both electromagnetic and gravitomagnetic potentials. For a neutral superconductor in the absence of external electromagnetic fields, the phase evolution along a worldline is governed by \cite{Papini:1967,Anandan:1977}
\begin{equation}\label{eq:phase_evol}
\frac{d\theta}{d\tau} = -\frac{m^*}{\hbar}g_{\mu\nu}u^\mu u^\nu - \frac{m^*}{\hbar}g_{t\phi}\frac{d\phi}{d\tau},
\end{equation}
where $u^\mu = dx^\mu/d\tau$ is the four-velocity and $\tau$ is the proper time along the worldline. The first term represents the standard phase evolution proportional to the invariant interval, while the second term captures the gravitomagnetic coupling through the off-diagonal metric component $g_{t\phi}$.

For a stationary interferometer with components at rest in the BL frame, the four-velocity has only a temporal component: $u^\mu = (u^t, 0, 0, 0)$. The normalization condition $g_{\mu\nu}u^\mu u^\nu = -1$ yields
\begin{equation}\label{eq:ut_norm}
u^t = \frac{1}{\sqrt{-g_{tt}}} = \frac{1}{\sqrt{1 - 2Mr/\Sigma}}.
\end{equation}
The phase accumulated over proper time $\tau$ along a path from point A to point B is then
\begin{equation}\label{eq:phase_path}
\theta_B - \theta_A = -\frac{m^*}{\hbar}\int_A^B \sqrt{-g_{\mu\nu}dx^\mu dx^\nu}.
\end{equation}

When the path includes azimuthal motion around the BH, the gravitomagnetic contribution becomes significant. Consider a path element with $dr = d\vartheta = 0$ but $d\phi \neq 0$. The line element gives
\begin{equation}\label{eq:ds_azimuthal}
ds^2 = g_{tt}dt^2 + 2g_{t\phi}dt\,d\phi + g_{\phi\phi}d\phi^2.
\end{equation}
For an observer at rest ($d\phi/dt = 0$), the phase evolution receives a contribution from $g_{t\phi}$ through the frame-dragging effect.

The physical interpretation becomes clearer in the ZAMO frame. A ZAMO rotates with angular velocity $\Omega_{\text{ZAMO}} = -g_{t\phi}/g_{\phi\phi}$ relative to distant stars, and in this locally non-rotating frame, the Cooper pair experiences an effective gravitomagnetic potential. The phase shift arises because different paths through the Kerr geometry sample different values of this potential.

\subsection{Gauge Invariance}

The gravitomagnetic phase shift for a closed loop $\mathcal{C}$ is gauge-invariant, depending only on the enclosed gravitomagnetic flux. Integrating the phase gradient around the loop yields
\begin{equation}\label{eq:phase_closed}
\Delta\theta = \oint_{\mathcal{C}} d\theta = \frac{m^*}{\hbar}\oint_{\mathcal{C}} g_{t\phi}\,d\phi.
\end{equation}
This expression is directly analogous to the electromagnetic AB phase
\begin{equation}\label{eq:AB_EM}
\Delta\theta_{\text{EM}} = \frac{e^*}{\hbar}\oint_{\mathcal{C}} A_\mu\, dx^\mu,
\end{equation}
with the gravitomagnetic potential $g_{t\phi}$ playing the role of the vector potential component and the mass $m^*$ replacing the charge $e^*$.

The gauge invariance of Eq.~\eqref{eq:phase_closed} can be understood as follows. Under a coordinate transformation $\phi \to \phi' = \phi + f(t,r,\vartheta)$, the metric component $g_{t\phi}$ transforms, but the integral $\oint g_{t\phi}\,d\phi$ around a closed loop remains unchanged because the transformation is single-valued. This mirrors the electromagnetic case, where the AB phase depends only on the magnetic flux through the loop, not on the choice of gauge for the vector potential.

The gravitomagnetic flux through a surface $\mathcal{S}$ bounded by $\mathcal{C}$ can be written using Stokes' theorem as
\begin{equation}\label{eq:flux_stokes}
\Phi_g = \oint_{\mathcal{C}} g_{t\phi}\,d\phi = \int_{\mathcal{S}} \left(\nabla \times \mathbf{A}_g\right) \cdot d\mathbf{S},
\end{equation}
where the gravitomagnetic field $\mathbf{B}_g = \nabla \times \mathbf{A}_g$ plays the role of the magnetic field in the electromagnetic analogy. For the Kerr metric, this flux is non-zero whenever the loop encloses the rotation axis, reflecting the topological nature of the gravitomagnetic AB effect.

\subsection{Comparison with Electromagnetic AB Effect}

The structural parallel between the gravitomagnetic and electromagnetic AB effects clarifies the physical content of Eq.~\eqref{eq:phase_closed}. In the electromagnetic case, a solenoid carrying current produces magnetic flux $\Phi_B$ confined to its interior. Charged particles encircling the solenoid acquire a phase shift $\Delta\theta_{\text{EM}} = e^*\Phi_B/\hbar$ even though they travel through regions where $\mathbf{B} = 0$. The magnetic flux quantum $\Phi_0 = \pi\hbar/e = 2.07 \times 10^{-15}$ Wb produces a phase shift of exactly $2\pi$ radians for a Cooper pair.

In the gravitomagnetic case, the rotating BH plays the role of the solenoid. The angular momentum $J = Ma$ generates frame-dragging effects encoded in $g_{t\phi}$, and Cooper pairs encircling the BH acquire a phase shift proportional to the enclosed gravitomagnetic flux. Unlike the electromagnetic solenoid, the gravitomagnetic field cannot be perfectly confined, so the analogy is not exact. Nevertheless, the phase shift $\Delta\theta = (m^*/\hbar)\Phi_g$ remains well-defined and gauge-invariant.

The key difference lies in the coupling constant. The ratio of mass to charge for a Cooper pair is
\begin{equation}
\frac{m^*}{e^*} = \frac{2m_e}{2e} = 5.69 \times 10^{-12}~\text{kg/C}.
\end{equation}
This small ratio indicates that for comparable potential strengths, the gravitomagnetic phase would be suppressed relative to the electromagnetic phase. However, the gravitomagnetic potential from an astrophysical BH can be enormously larger than typical laboratory electromagnetic potentials. For a $10\,M_\odot$ BH with $a/M = 0.9$ and a circular path at $r = 3M$, the gravitomagnetic flux reaches $|\Phi_g| \approx 5.57 \times 10^4$ m in SI units. The resulting phase shift, which we calculate in detail in Sec.~\ref{sec4}, can be substantial for paths close to the horizon. The phase scales linearly with BH mass, reaching even larger values for supermassive BHs.

Table~\ref{tab:AB_comparison} summarizes the correspondence between the two phenomena, while Table~\ref{tab:numerical_comparison} provides a direct numerical comparison of the relevant parameters.

\begin{table}[ht!]
\centering
\caption{Comparison between electromagnetic and gravitomagnetic AB effects for Cooper pairs \cite{Aharonov:1959,DeWitt:1966,Anandan:1977}. The structural similarity enables direct translation of experimental concepts from superconducting interferometry to gravitational contexts.}
\label{tab:AB_comparison}
\begin{tabular}{c|c|c}
\hline\hline
\rowcolor{orange!50}
Quantity & Electromagnetic AB & Gravitomagnetic AB \\
\hline
Coupling constant & $e^* = 3.20 \times 10^{-19}$ C & $m^* = 1.82 \times 10^{-30}$ kg \\
Vector potential & $A_\mu$ & $g_{t\phi}$ \\
Field strength & $\mathbf{B} = \nabla \times \mathbf{A}$ & $\mathbf{B}_g = \nabla \times \mathbf{A}_g$ \\
Flux quantum & $\Phi_0 = 2.07 \times 10^{-15}$ Wb & --- \\
Phase per flux quantum & $2\pi$ rad & --- \\
Source & Solenoid current & BH angular momentum $J$ \\
\hline\hline
\end{tabular}
\end{table}

\begin{table}[ht!]
\centering
\caption{Numerical comparison of electromagnetic and gravitomagnetic AB parameters for Cooper pairs \cite{Tinkham:2004,Mashhoon:2003}. The ratio $m^*/e^* = 5.69 \times 10^{-12}$ kg/C determines the relative sensitivity of mass versus charge coupling.}
\label{tab:numerical_comparison}
\begin{tabular}{c|c|c}
\hline\hline
\rowcolor{orange!50}
Parameter & Electromagnetic & Gravitomagnetic \\
\hline
Coupling constant & $e^* = 3.20 \times 10^{-19}$ C & $m^* = 1.82 \times 10^{-30}$ kg \\
Typical potential & $A \sim 10^{-6}$ Wb/m & $g_{t\phi} \sim 10^{4}$ m (for $10^6\,M_\odot$) \\
Phase for reference flux & $2\pi$ rad per $\Phi_0$ & $\sim 10^{17}$ rad (for $10\,M_\odot$, $r=3M$) \\
\hline\hline
\end{tabular}
\end{table}

\subsection{Coherence Requirements}

For the gravitomagnetic AB effect to be observable, the superconducting condensate must maintain phase coherence over the entire interferometer loop. The coherence length $\xi$ and the London penetration depth $\lambda_L$ characterize the length scales over which the order parameter and the supercurrent vary, respectively \cite{Tinkham:2004}.

In a type-II superconductor, the GL coherence length is given by
\begin{equation}\label{eq:coherence_length}
\xi(T) = \xi_0 \sqrt{\frac{T_c}{T_c - T}},
\end{equation}
where $\xi_0$ is the zero-temperature coherence length and $T_c$ is the critical temperature. Table~\ref{tab:coherence} lists the coherence lengths for representative superconducting materials. Conventional superconductors like aluminum offer the longest coherence lengths, making them preferable for interferometric applications.

\begin{table}[ht!]
\centering
\caption{Zero-temperature coherence lengths $\xi_0$ for representative 
superconducting materials \cite{Tinkham:2004,Poole:2007}. Longer coherence 
lengths facilitate larger interferometer loop sizes.}
\label{tab:coherence}
\begin{tabular}{c|c|c}
\hline\hline
\rowcolor{orange!50}
Material & Type & $\xi_0$ \\
\hline
Aluminum (Al) & Type-I & $1.6~\mu$m \\
Niobium (Nb) & Type-II & $38$ nm \\
YBCO (high-$T_c$) & Type-II & $1.5$ nm \\
\hline\hline
\end{tabular}
\end{table}

The requirement for observing quantum interference is that the path length difference between the two arms of the interferometer must be smaller than the phase coherence length. In SQUID devices, this condition is readily satisfied for loop sizes up to millimeters or even centimeters at sufficiently low temperatures.

In the gravitational context, an additional consideration arises: tidal forces from the BH can disrupt the superconducting state. The tidal acceleration across a Cooper pair of size $\xi$ at distance $r$ from a BH of mass $M$ is approximately
\begin{equation}\label{eq:tidal}
a_{\text{tidal}} \sim \frac{GM\xi}{r^3}.
\end{equation}
For a supermassive BH with $M = 10^6\,M_\odot$ at $r = 10\,r_s$ (where $r_{\text{dist}} = 2.95 \times 10^{10}$ m) and $\xi = 1~\mu$m, the tidal acceleration is
\begin{equation}
a_{\text{tidal}} = 5.15 \times 10^{-12}~\text{m/s}^2.
\end{equation}
This should be compared with the characteristic acceleration scale set by the Cooper pair binding energy $E_{\text{bind}} \sim k_B T_c$:
\begin{equation}
a_{\text{bind}} \sim \frac{E_{\text{bind}}}{m^* \xi} = \frac{k_B T_c}{m^* \xi} = 7.58 \times 10^{13}~\text{m/s}^2,
\end{equation}
where we used $T_c = 10$ K and $k_B = 1.38 \times 10^{-23}$ J/K. The ratio
\begin{equation}
\frac{a_{\text{tidal}}}{a_{\text{bind}}} = 6.79 \times 10^{-26}
\end{equation}
confirms that tidal disruption of Cooper pairs is completely negligible at distances $r \gg r_+$ where our analysis applies.

The practical conclusion is that superconducting coherence can in principle be maintained at astrophysically relevant distances from BHs, though the phase shifts become correspondingly smaller at large distances due to the $1/r$ falloff of $g_{t\phi}$.

{\color{black}

\section{AB Phase Calculation} \label{sec4}

With the theoretical framework established in the preceding sections, we now compute the gravitomagnetic AB phase for specific interferometer configurations in Kerr spacetime. The phase depends on the BH parameters $(M, a)$, the interferometer geometry $(r_1, r_2, \iota)$, and the path topology. We derive closed-form expressions for equatorial and inclined orbits, analyze the parameter dependence, and identify optimal configurations that maximize the phase signal. The calculations reveal that while the gravitomagnetic coupling is intrinsically weak, the enormous scale of the gravitomagnetic potential near astrophysical BHs can produce substantial phase shifts that encode information about the BH spin and mass.

\subsection{Equatorial Interferometer}

Consider a Cooper pair interferometer in the equatorial plane with two arms encircling the BH at different radii $r_1$ and $r_2$, where $r_2 > r_1 > r_+$. The interferometer operates by splitting a Cooper pair beam, sending the two components along circular paths at the respective radii, and recombining them after a complete azimuthal circuit. The phase difference between the two arms arises from the different values of the gravitomagnetic potential sampled by each path.

Using Eq.~\eqref{eq:phase_closed} from Sec.~\ref{sec3}, the phase accumulated by a Cooper pair traversing a complete circular path at radius $r$ in the equatorial plane is
\begin{equation}\label{eq:phase_single}
\theta(r) = \frac{m^*}{\hbar}\oint g_{t\phi}\big|_{\vartheta=\pi/2} d\phi = \frac{m^*}{\hbar} \cdot \left(-\frac{2Ma}{r}\right) \cdot 2\pi = -\frac{4\pi m^* Ma}{\hbar r}.
\end{equation}
The phase difference between the two interferometer arms is therefore
\begin{equation}\label{eq:AB_equatorial}
\Delta\theta_{\text{AB}}^{(\text{eq})} = \theta(r_1) - \theta(r_2) = \frac{4\pi m^* Ma}{\hbar}\left(\frac{1}{r_2} - \frac{1}{r_1}\right).
\end{equation}
Note the sign: since $r_1 < r_2$, we have $1/r_1 > 1/r_2$, making the quantity in parentheses negative. The inner arm accumulates a larger (more negative) phase due to stronger frame dragging closer to the BH.

Restoring SI units, the gravitomagnetic flux involves factors of $G$ and $c$. The spin parameter in SI units is $a_{\text{SI}} = (a/M) \cdot GM/c^2$, and the radii are related to the Schwarzschild radius $r_s = 2GM/c^2$ by $r = (r/r_s) \cdot r_s$. The phase difference becomes
\begin{equation}\label{eq:AB_SI}
\Delta\theta_{\text{AB}}^{(\text{eq})} = \frac{4\pi G m^* M (a/M)}{\hbar c^3}\left(\frac{1}{r_2} - \frac{1}{r_1}\right) = \frac{4\pi G m^* J}{\hbar c^3}\left(\frac{1}{r_2} - \frac{1}{r_1}\right),
\end{equation}
where $J = Ma$ is the BH angular momentum. This expression makes manifest that the phase is directly proportional to $J$, providing a quantum-coherent measurement of the BH spin.

The phase can also be written in terms of the Schwarzschild radius:
\begin{equation}\label{eq:AB_rs}
\Delta\theta_{\text{AB}}^{(\text{eq})} = \frac{2\pi m^* c}{\hbar} \cdot (a/M) \cdot r_s \left(\frac{1}{r_2} - \frac{1}{r_1}\right).
\end{equation}
For radii expressed in units of $r_s$, with $\tilde{r} = r/r_s$, this becomes
\begin{equation}\label{eq:AB_dimless}
\Delta\theta_{\text{AB}}^{(\text{eq})} = \frac{2\pi m^* c}{\hbar} \cdot (a/M) \left(\frac{1}{\tilde{r}_2} - \frac{1}{\tilde{r}_1}\right).
\end{equation}

The prefactor $m^* c/\hbar = 2m_e c/\hbar = 5.17 \times 10^{12}$ m$^{-1}$ sets the scale of the phase per unit gravitomagnetic potential. For a $10\,M_\odot$ BH with $a/M = 0.9$ and interferometer arms at $\tilde{r}_1 = 1.5$ and $\tilde{r}_2 = 10$, the phase difference is
\begin{equation}
\Delta\theta_{\text{AB}}^{(\text{eq})} \approx 2.9 \times 10^{17}~\text{rad}.
\end{equation}
This enormous phase reflects the accumulation over the $2\pi$ azimuthal circuit near a strongly gravitating object. While measuring such a large phase directly is impractical, the physically relevant quantity is the phase modulo $2\pi$, or equivalently, the interference fringe pattern that results from the phase difference.

\subsection{Dependence on BH Parameters}

The AB phase exhibits a linear dependence on BH mass $M$ (at fixed dimensionless radii $r/r_s$) and spin parameter $a/M$. This scaling has important implications for using the gravitomagnetic AB effect as a probe of BH properties.

From Eq.~\eqref{eq:AB_SI}, we see that at fixed dimensionless radii $\tilde{r}_1 = r_1/r_s$ and $\tilde{r}_2 = r_2/r_s$:
\begin{equation}\label{eq:scaling}
\Delta\theta = \frac{2\pi G m^* M (a/M)}{\hbar c}\left(\frac{1}{\tilde{r}_1} - \frac{1}{\tilde{r}_2}\right) \propto M \cdot (a/M).
\end{equation}
The phase is directly proportional to the BH angular momentum $J = Ma$. This linear scaling means that supermassive BHs produce proportionally larger phases than stellar-mass BHs, making them more favorable targets despite their greater distances.

Figure~\ref{fig:phase_vs_r1} illustrates the dependence of the AB phase on the inner arm radius $r_1$ for a supermassive BH with $M = 10^6\,M_\odot$ and spin parameter $a/M = 0.9$. The outer arm is fixed at $r_2 = 10\,r_s$. As the inner arm approaches the horizon (marked by the vertical dashed line at $r_+ \approx 1.44\,r_s$), the phase increases dramatically, reaching $|\Delta\theta| \sim 10^{22.5}$ rad near the horizon. A characteristic dip appears at $r_1 = r_2 = 10\,r_s$, where the enclosed gravitomagnetic flux between the arms vanishes and the phase difference goes to zero. For $r_1 > r_2$, the phase changes sign but its magnitude increases again, corresponding to configurations where the ``inner'' arm is actually exterior to the ``outer'' arm. This behavior underscores the topological nature of the AB effect: the phase depends on the \textit{enclosed} gravitomagnetic flux between the two arms, which vanishes when $r_1 = r_2$ and changes sign when their ordering reverses.

\begin{figure}[ht!]
\centering
\includegraphics[width=0.85\textwidth]{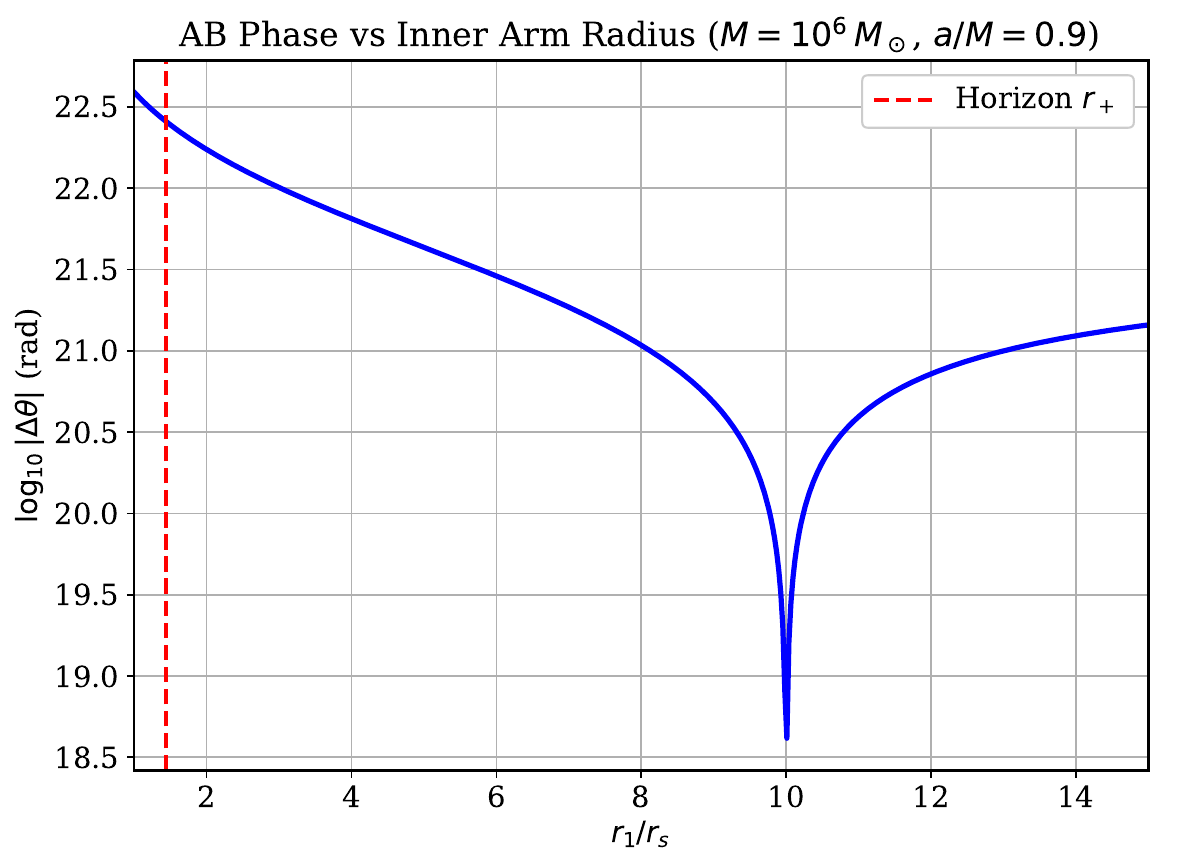}
\caption{Gravitomagnetic AB phase as a function of the inner arm radius $r_1/r_s$ for a BH with $M = 10^6\,M_\odot$ and $a/M = 0.9$. The outer arm is fixed at $r_2 = 10\,r_s$. The vertical dashed line indicates the outer horizon at $r_+ = M(1 + \sqrt{1-(a/M)^2}) \approx 1.44\,r_s$. The phase diverges logarithmically as $r_1 \to r_+$ and exhibits a characteristic dip at $r_1 = r_2$ where the enclosed flux vanishes. The enormous magnitude ($\sim 10^{22}$ rad) reflects the macroscopic gravitomagnetic flux near astrophysical BHs.}
\label{fig:phase_vs_r1}
\end{figure}

The linear dependence of the AB phase on spin parameter is demonstrated in Fig.~\ref{fig:phase_vs_spin}. For fixed interferometer geometry ($r_1 = 2\,r_s$, $r_2 = 10\,r_s$) and BH mass ($M = 10^6\,M_\odot$), the phase scales as $|\Delta\theta| \propto a/M$, passing through the origin for a non-rotating (Schwarzschild) BH. This linear relationship is a direct consequence of the gravitomagnetic flux being proportional to the BH angular momentum $J = Ma$. The marked points correspond to moderate spin ($a/M = 0.5$), rapid spin characteristic of Sgr~A* ($a/M = 0.9$), and near-extremal rotation ($a/M = 0.99$). At $a/M = 0.9$, the phase reaches $|\Delta\theta| \approx 1.73 \times 10^{22}$ rad. This linear scaling implies that precise measurement of the AB phase would directly yield the BH spin parameter, providing a quantum-coherent probe of the Kerr geometry complementary to electromagnetic and gravitational-wave observations.

\begin{figure}[ht!]
\centering
\includegraphics[width=0.75\textwidth]{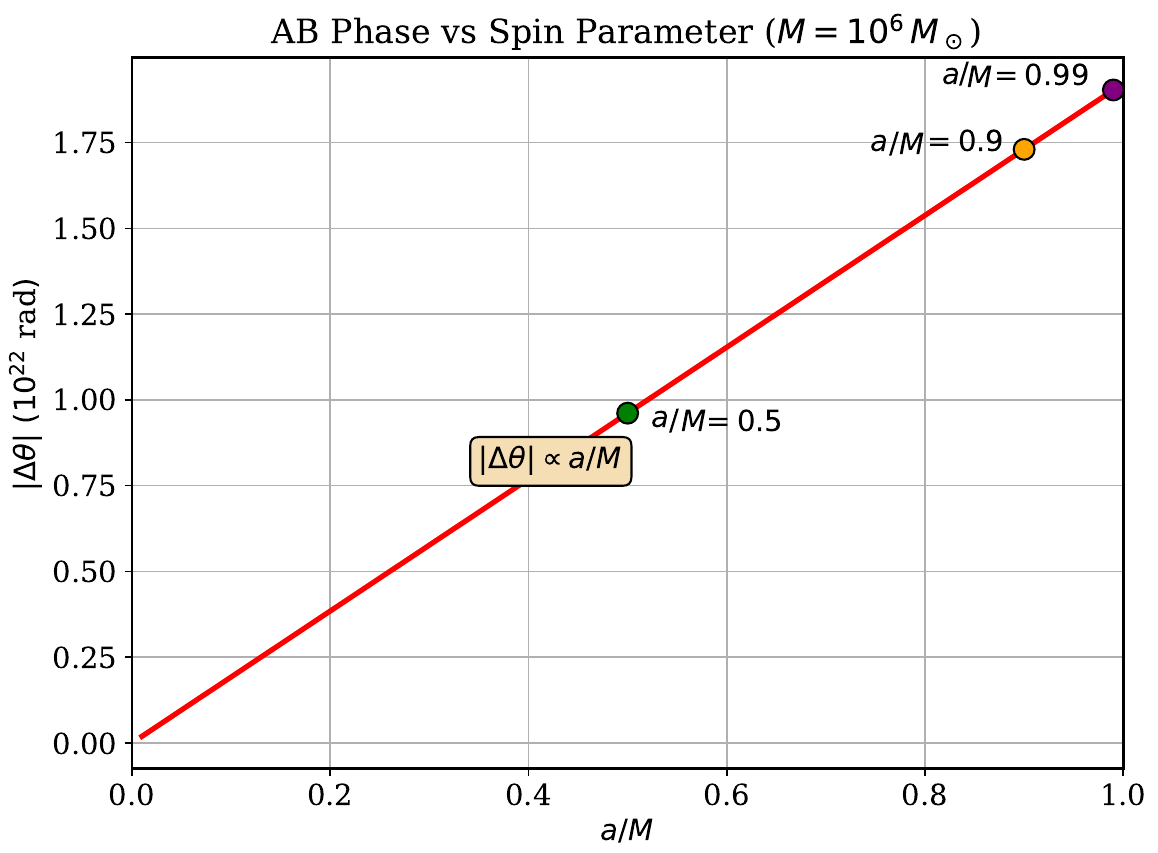}
\caption{AB phase versus dimensionless spin parameter $a/M$ for a BH with $M = 10^6\,M_\odot$ and interferometer arms at $r_1 = 2\,r_s$, $r_2 = 10\,r_s$. The phase exhibits strict linear scaling $|\Delta\theta| \propto a/M$, reflecting the proportionality of the gravitomagnetic flux to the BH angular momentum. Colored markers indicate $a/M = 0.5$ (green), $a/M = 0.9$ (orange), and $a/M = 0.99$ (purple). The linear relationship implies that the AB phase provides a direct quantum probe of BH spin.}
\label{fig:phase_vs_spin}
\end{figure}

Table~\ref{tab:phase_vs_mass} shows the AB phase for BHs spanning the astrophysical mass range, from stellar-mass objects ($M \sim 10\,M_\odot$) to the most massive known BHs ($M \sim 10^{9}\,M_\odot$). The phase increases by nine orders of magnitude across this range, tracking the linear mass dependence at fixed $r/r_s$.

\begin{table}[ht!]
\centering
\caption{Gravitomagnetic AB phase for BHs of various masses with $a/M = 0.9$, $r_1 = 2\,r_s$, and $r_2 = 10\,r_s$. The phase scales linearly with mass at fixed dimensionless radii.}
\label{tab:phase_vs_mass}
\begin{tabular}{c|c|c|c}
\hline\hline
\rowcolor{orange!50}
$M$ ($M_\odot$) & $r_s$ (km) & $|\Delta\theta|$ (rad) & $|\Delta\theta|$ (cycles) \\
\hline
$10$ & $29.5$ & $2.88 \times 10^{17}$ & $4.59 \times 10^{16}$ \\
$10^2$ & $295$ & $2.88 \times 10^{18}$ & $4.59 \times 10^{17}$ \\
$10^3$ & $2.95 \times 10^3$ & $2.88 \times 10^{19}$ & $4.59 \times 10^{18}$ \\
$10^4$ & $2.95 \times 10^4$ & $2.88 \times 10^{20}$ & $4.59 \times 10^{19}$ \\
$10^5$ & $2.95 \times 10^5$ & $2.88 \times 10^{21}$ & $4.59 \times 10^{20}$ \\
$10^6$ & $2.95 \times 10^6$ & $2.88 \times 10^{22}$ & $4.59 \times 10^{21}$ \\
$10^7$ & $2.95 \times 10^7$ & $2.88 \times 10^{23}$ & $4.59 \times 10^{22}$ \\
$10^8$ & $2.95 \times 10^8$ & $2.88 \times 10^{24}$ & $4.59 \times 10^{23}$ \\
$10^9$ & $2.95 \times 10^9$ & $2.88 \times 10^{25}$ & $4.59 \times 10^{24}$ \\
\hline\hline
\end{tabular}
\end{table}

Figure~\ref{fig:phase_vs_mass} presents the AB phase as a function of BH mass on a log-log scale, spanning from stellar-mass BHs ($M \sim 10\,M_\odot$) to the most massive SMBHs ($M \sim 10^{10}\,M_\odot$). With the interferometer arms positioned at fixed multiples of the Schwarzschild radius ($r_1 = 2\,r_s$, $r_2 = 10\,r_s$) and spin $a/M = 0.9$, the phase scales linearly with mass: $|\Delta\theta| \propto M$. This linear scaling arises because, while the phase formula contains $M^2$ in the numerator, the geometric factor $(1/r_1 - 1/r_2)$ scales as $1/r_s \propto 1/M$ when the arm positions are defined as fixed multiples of $r_s$. The product $M^2 \times M^{-1} = M$ thus yields unit slope on the log-log plot. Three astrophysical reference points are marked: a typical stellar-mass BH ($10\,M_\odot$, $|\Delta\theta| \sim 10^{17}$~rad), Sgr~A* ($4.3 \times 10^6\,M_\odot$, $|\Delta\theta| \sim 10^{23}$~rad), and M87* ($6.5 \times 10^9\,M_\odot$, $|\Delta\theta| \sim 10^{26}$~rad). The nine orders of magnitude span in phase between stellar and supermassive BHs reflects the corresponding range in gravitomagnetic flux, with SMBHs providing by far the strongest signal for any hypothetical AB interferometer.
\begin{figure}[ht!]
\centering
\includegraphics[width=0.75\textwidth]{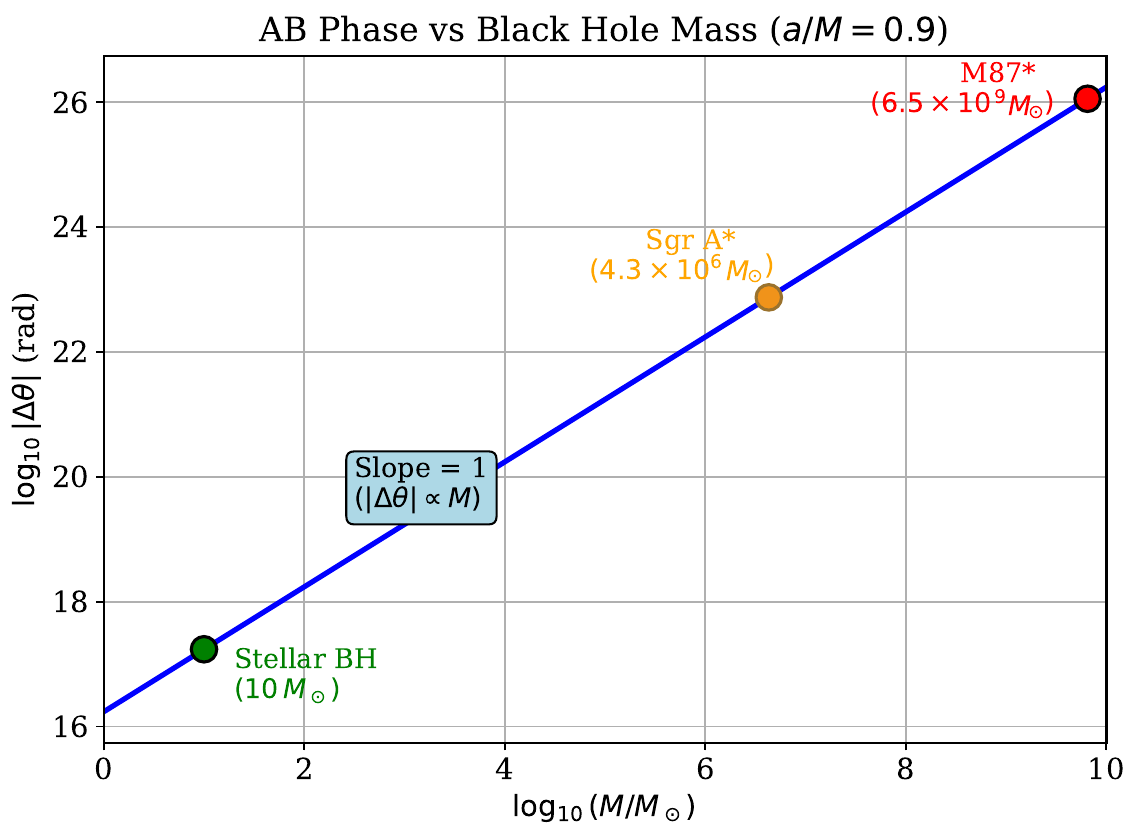}
\caption{Log-log plot of the AB phase versus BH mass for $a/M = 0.9$ with arms at $r_1 = 2\,r_s$ and $r_2 = 10\,r_s$. The unit slope confirms the linear scaling $|\Delta\theta| \propto M$. Three astrophysical systems are marked: a stellar-mass BH ($10\,M_\odot$, green), Sgr~A* ($4.3 \times 10^6\,M_\odot$, orange), and M87* ($6.5 \times 10^9\,M_\odot$, red). The predicted phases span from $\sim 10^{17}$ rad for stellar BHs to $\sim 10^{26}$ rad for M87*, demonstrating that supermassive BHs generate enormously larger gravitomagnetic AB effects.}
\label{fig:phase_vs_mass}
\end{figure}

\subsection{Inclined Orbits}

For an interferometer whose orbital plane is tilted at angle $\iota$ relative to the BH equatorial plane, the gravitomagnetic flux is reduced. The physical reason is that $g_{t\phi} \propto \sin^2\vartheta$, so paths passing through higher latitudes (smaller $\sin\vartheta$) sample weaker gravitomagnetic potentials.

For an inclined circular orbit, the polar angle $\vartheta$ varies with azimuthal angle $\phi$ according to the geometry of the tilted plane. In the limit of small spin ($a \ll r$) and moderate inclinations, the path-averaged gravitomagnetic potential is reduced by a factor of approximately $\cos\iota$:
\begin{equation}\label{eq:AB_inclined}
\Delta\theta_{\text{AB}}^{(\iota)} \approx \Delta\theta_{\text{AB}}^{(\text{eq})} \cos\iota.
\end{equation}
This approximation captures the leading-order inclination dependence and becomes exact in the Newtonian (weak-field) limit.

The full calculation requires integrating $g_{t\phi}(\vartheta(\phi))$ over the inclined path. For an orbit inclined at angle $\iota$, the polar angle varies as
\begin{equation}\label{eq:theta_phi}
\cos\vartheta = \sin\iota \sin\phi,
\end{equation}
where $\phi$ is measured from the ascending node. Substituting into $g_{t\phi} = -2Mar\sin^2\vartheta/\Sigma$ and integrating over $\phi \in [0, 2\pi]$ yields corrections to Eq.~\eqref{eq:AB_inclined} of order $(a/r)^2$.

The reduction of the AB phase for inclined interferometer configurations is shown in Fig.~\ref{fig:phase_vs_incl}. Panel~(a) displays the normalized phase $|\Delta\theta(\iota)|/|\Delta\theta_{\rm eq}|$ as a function of inclination angle $\iota$, measured from the equatorial plane. The cosine dependence, $|\Delta\theta(\iota)| = |\Delta\theta_{\rm eq}|\cos\iota$, arises from the projection of the gravitomagnetic flux onto the inclined orbital plane. At $\iota = 30°$, $45°$, and $60°$, the phase is reduced to $87\%$, $71\%$, and $50\%$ of its equatorial value, respectively. For polar orbits ($\iota = 90°$), the phase vanishes entirely as the orbit encloses no net gravitomagnetic flux. Panel~(b) shows the corresponding absolute phase for $M = 10^6\,M_\odot$, with the equatorial value $|\Delta\theta_{\rm eq}| \approx 1.73 \times 10^{22}$ rad. These results indicate that equatorial configurations are optimal for maximizing the gravitomagnetic signal, though substantial phases remain accessible even for moderately inclined orbits.

\begin{figure}[ht!]
\centering
\includegraphics[width=0.95\textwidth]{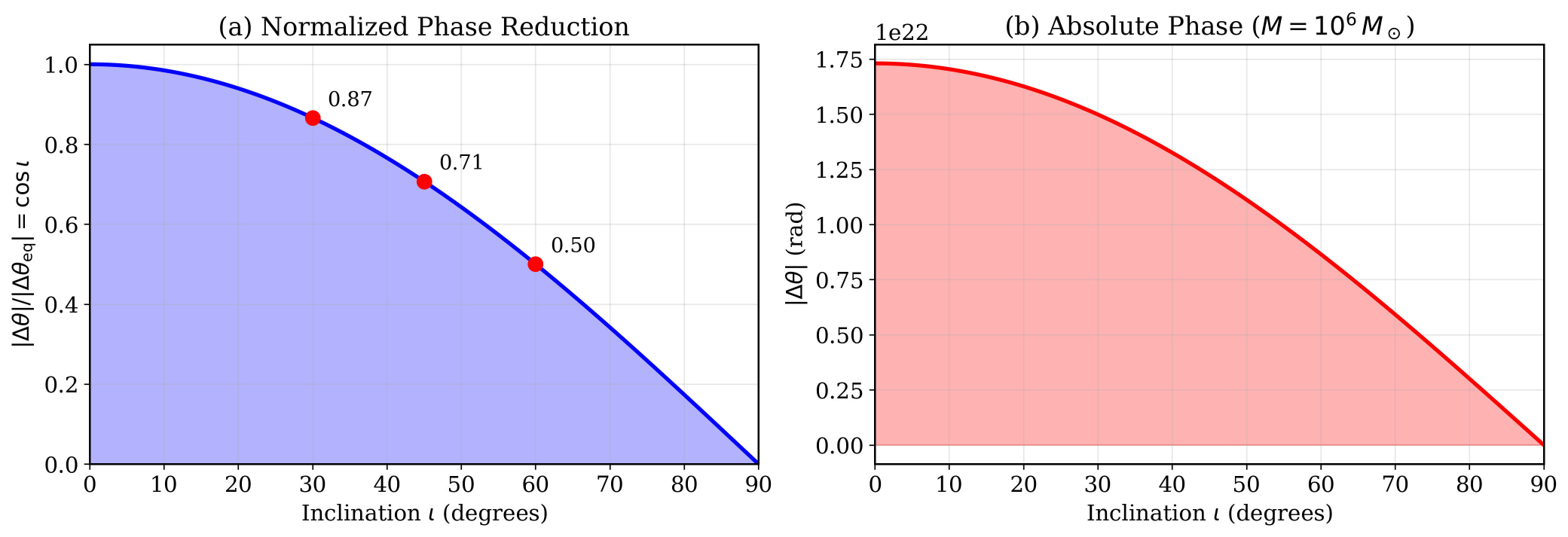}
\caption{Dependence of the AB phase on orbital inclination angle $\iota$. (a)~Normalized phase reduction factor $|\Delta\theta|/|\Delta\theta_{\rm eq}| = \cos\iota$, with markers at $\iota = 30°$ (0.87), $45°$ (0.71), and $60°$ (0.50). (b)~Absolute phase for $M = 10^6\,M_\odot$, $a/M = 0.9$, showing the equatorial maximum of $1.73 \times 10^{22}$ rad decreasing to zero at polar inclination. The cosine reduction reflects the projection of the enclosed gravitomagnetic flux onto the inclined orbital plane.}
\label{fig:phase_vs_incl}
\end{figure}

Table~\ref{tab:inclination} summarizes the reduction factor as a function of inclination angle. This strong inclination dependence means that equatorial configurations are essential for maximizing the signal.

\begin{table}[htbp]
\centering
\caption{Inclination reduction factor $\cos\iota$ for the gravitomagnetic AB phase. Equatorial orbits ($\iota = 0$) maximize the phase.}
\label{tab:inclination}
\begin{tabular}{c|c|c}
\hline\hline
\rowcolor{orange!50}
$\iota$ (degrees) & $\cos\iota$ & Fraction of equatorial phase \\
\hline
$0$ & $1.000$ & $100\%$ \\
$15$ & $0.966$ & $96.6\%$ \\
$30$ & $0.866$ & $86.6\%$ \\
$45$ & $0.707$ & $70.7\%$ \\
$60$ & $0.500$ & $50.0\%$ \\
$75$ & $0.259$ & $25.9\%$ \\
$90$ & $0.000$ & $0\%$ \\
\hline\hline
\end{tabular}
\end{table}

\subsection{Optimal Configuration and Arm Separation}

The gravitomagnetic AB phase is maximized by choosing the interferometer geometry to sample the strongest frame-dragging regions while maintaining experimental feasibility. Several factors contribute to optimizing the signal.

Equatorial orientation ($\iota = 0$) is essential, as any inclination reduces the phase by the factor $\cos\iota$. The inner arm should be placed as close to the horizon as possible, since $g_{t\phi} \propto 1/r$ near the equator implies that the phase contribution diverges as $r_1 \to r_+$. However, practical considerations such as tidal stability and radiation environment limit how close to the horizon the interferometer can operate. A reasonable minimum might be $r_1 \sim 1.5\,r_s$ for rapidly rotating BHs or $r_1 \sim 3\,r_s$ for safety margins. The outer arm radius $r_2$ should be large enough to capture most of the gravitomagnetic flux, but the returns diminish as $r_2$ increases since the potential falls as $1/r$. Beyond $r_2 \sim 10$--$20\,r_s$, further increases contribute little to the phase difference. Rapid BH rotation ($a \to M$) is favorable since the phase scales linearly with $a$. Observed astrophysical BHs span a range of spins, with some approaching the extremal limit $a/M \sim 0.99$.

The full parameter space of interferometer configurations is explored in Fig.~\ref{fig:phase_contour}, which presents the AB phase as a function of both arm radii $(r_1, r_2)$ for a BH with $M = 10^6\,M_\odot$ and $a/M = 0.9$. The color scale represents $\log_{10}|\Delta\theta|$, with warmer colors indicating larger phases. The physically relevant region lies above the diagonal $r_2 = r_1$ (white dashed line), where the outer arm is indeed exterior to the inner arm. The phase is maximized in the upper-left corner of the parameter space, corresponding to configurations with $r_1$ close to the horizon and $r_2$ far from the BH. Along the diagonal, the phase approaches its minimum (dark region) as the enclosed gravitomagnetic flux between the arms vanishes. The cyan star marks our reference configuration ($r_1 = 2\,r_s$, $r_2 = 10\,r_s$), which yields $|\Delta\theta| \approx 10^{22}$ rad. This contour map provides a practical guide for optimizing interferometer geometry: configurations with small $r_1$ and large $r_2$ maximize the gravitomagnetic signal, though the inner arm must remain outside the horizon for physical accessibility.

\begin{figure}[ht!]
\centering
\includegraphics[width=0.75\textwidth]{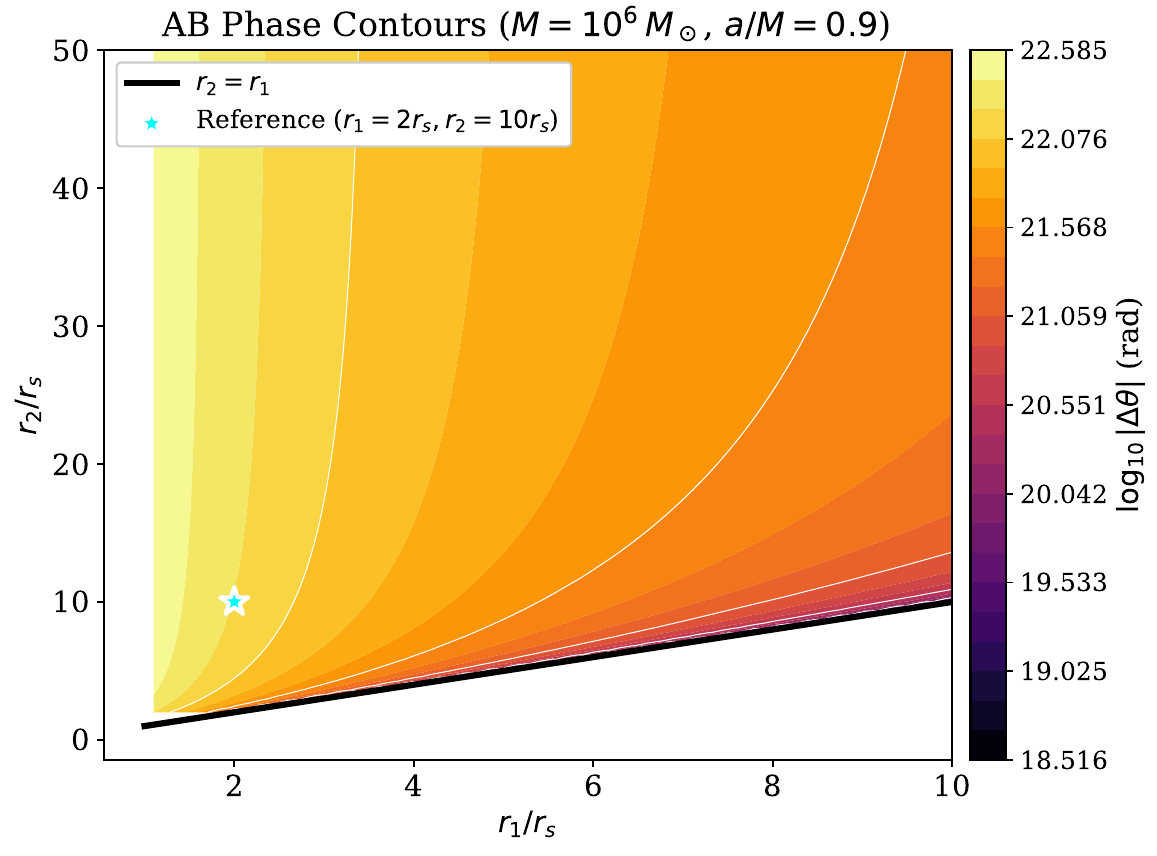}
\caption{Contour plot of $\log_{10}|\Delta\theta|$ in the $(r_1/r_s, r_2/r_s)$ parameter plane for $M = 10^6\,M_\odot$ and $a/M = 0.9$. The black solid line indicates $r_2 = r_1$, below which the configuration is unphysical (outer arm interior to inner arm). The phase is maximized for small $r_1$ and large $r_2$ (upper-left, yellow region) and minimized along the diagonal where the enclosed flux vanishes (dark region). The cyan star marks the reference configuration $(r_1, r_2) = (2\,r_s, 10\,r_s)$. This map guides optimal interferometer design for maximizing the gravitomagnetic AB signal.}
\label{fig:phase_contour}
\end{figure}

For an extremal Kerr BH ($a = M$) with $r_1 = 2M$ (equivalently $r_1 = r_s$) and $r_2 = 10M$ ($r_2 = 5\,r_s$), the geometric factor becomes
\begin{equation}\label{eq:geom_factor}
\frac{1}{r_2} - \frac{1}{r_1} = \frac{1}{10M} - \frac{1}{2M} = -\frac{4}{10M} = -\frac{2}{5M}.
\end{equation}
The phase difference is then
\begin{equation}\label{eq:AB_extremal}
\Delta\theta_{\text{AB}} = \frac{4\pi m^* M}{\hbar} \cdot \left(-\frac{2}{5M}\right) = -\frac{8\pi m^*}{5\hbar}.
\end{equation}
In SI units, this evaluates to $|\Delta\theta_{\text{AB}}| = 8\pi m^* c/(5\hbar) \approx 8.6 \times 10^{12}$ radians per meter of $r_s$, or equivalently per $(GM/c^2)$.

The interferometric phase difference depends on the arm separation through the factor $(1/r_1 - 1/r_2)$. This factor has a simple geometric interpretation: it measures the difference in solid angle subtended by the two circular arms as seen from the BH. The gravitomagnetic flux enclosed between the arms determines the phase, in direct analogy with the magnetic flux threading a SQUID loop in conventional superconducting interferometry \cite{Jaklevic:1964,Clarke:2004}.

The connection to flux can be made explicit by applying Stokes' theorem to the gravitomagnetic potential. The phase accumulated around a closed loop $\mathcal{C}$ is
\begin{equation}\label{eq:stokes_phase}
\Delta\theta = \frac{m^*}{\hbar}\oint_{\mathcal{C}} g_{t\phi}\,d\phi = \frac{m^*}{\hbar}\int_{\mathcal{S}} (\nabla \times \mathbf{A}_g) \cdot d\mathbf{S},
\end{equation}
where $\mathbf{A}_g$ is the gravitomagnetic vector potential and $\mathcal{S}$ is any surface bounded by $\mathcal{C}$ \cite{Mashhoon:2003,Ruggiero:2002}. For an annular interferometer with inner radius $r_1$ and outer radius $r_2$, the enclosed gravitomagnetic flux is
\begin{equation}\label{eq:gm_flux_enclosed}
\Phi_g^{\text{enc}} = \int_{r_1}^{r_2} \int_0^{2\pi} B_g^z \, r\,dr\,d\phi,
\end{equation}
where $B_g^z = (\nabla \times \mathbf{A}_g)_z$ is the gravitomagnetic field component normal to the equatorial plane. In the weak-field limit, $B_g^z \approx 2GJ/(c^2 r^3)$ \cite{Thorne:1986}, giving
\begin{equation}\label{eq:flux_integral}
\Phi_g^{\text{enc}} = \frac{4\pi GJ}{c^2}\left(\frac{1}{r_1} - \frac{1}{r_2}\right),
\end{equation}
which recovers the radial dependence in Eq.~\eqref{eq:AB_equatorial}.

This flux-based interpretation highlights the topological nature of the AB effect: the phase depends only on the total flux enclosed, not on the detailed path taken by the Cooper pairs. This is analogous to the electromagnetic AB effect, where electrons acquire a phase proportional to the magnetic flux through their path, even in regions where $\mathbf{B} = 0$ \cite{Aharonov:1959,Peshkin:1989}. The key difference is that while magnetic flux can be confined to a solenoid, the gravitomagnetic field of a Kerr BH extends throughout spacetime, decaying as $1/r^3$ in the weak-field regime.

For fixed outer radius $r_2$, the phase increases as $r_1$ decreases toward the horizon. The rate of increase is largest near the horizon:
\begin{equation}\label{eq:dphase_dr1}
\frac{\partial|\Delta\theta|}{\partial r_1} = \frac{4\pi m^* Ma}{\hbar r_1^2},
\end{equation}
which diverges as $r_1 \to 0$. This strong sensitivity to the inner radius motivates placing the inner arm as close as feasible to the horizon. The divergence is ultimately regulated by the horizon radius $r_+$, where the coordinate system breaks down and the physical interpretation must be modified \cite{Bardeen:1972}.

Conversely, for fixed inner radius $r_1$, the phase saturates as $r_2 \to \infty$:
\begin{equation}\label{eq:phase_saturation}
\lim_{r_2 \to \infty} \Delta\theta_{\text{AB}}^{(\text{eq})} = -\frac{4\pi m^* Ma}{\hbar r_1}.
\end{equation}
This saturation reflects that most of the gravitomagnetic flux is concentrated near the BH; the outer regions contribute little to the total. In practice, $r_2 \gtrsim 10\,r_s$ captures more than 90\% of the asymptotic phase for $r_1 \sim 2\,r_s$.

Table~\ref{tab:arm_separation} shows the phase for various arm configurations, illustrating the diminishing returns from increasing $r_2$ and the strong gains from decreasing $r_1$.

\begin{table}[htbp]
\centering
\caption{AB phase dependence on arm separation for $M = 10^6\,M_\odot$ and $a/M = 0.9$ \cite{Mashhoon:2003}. Decreasing $r_1$ toward the horizon yields the largest gains, while increasing $r_2$ beyond $\sim 10\,r_s$ provides diminishing returns.}
\label{tab:arm_separation}
\begin{tabular}{c|c|c|c}
\hline\hline
\rowcolor{orange!50}
$r_1/r_s$ & $r_2/r_s$ & $\Delta r/r_s$ & $|\Delta\theta|$ (rad) \\
\hline
$1.0$ & $10$ & $9.0$ & $2.59 \times 10^{22}$ \\
$2.0$ & $10$ & $8.0$ & $1.15 \times 10^{22}$ \\
$3.0$ & $10$ & $7.0$ & $6.74 \times 10^{21}$ \\
$2.0$ & $20$ & $18$ & $1.30 \times 10^{22}$ \\
$2.0$ & $50$ & $48$ & $1.38 \times 10^{22}$ \\
$2.0$ & $100$ & $98$ & $1.41 \times 10^{22}$ \\
\hline\hline
\end{tabular}
\end{table}

\subsection{Astrophysical BH Scenarios}

To connect the theoretical calculations to observable systems, we evaluate the AB phase for specific astrophysical BHs with known or estimated parameters.

\textit{Sagittarius A*}: The supermassive BH at the center of the Milky Way has mass $M = (4.3 \pm 0.01) \times 10^6\,M_\odot$ and Schwarzschild radius $r_s = 1.27 \times 10^7$ km, corresponding to about $0.085$ AU or $42$ light-seconds. Spin measurements from stellar dynamics and flare observations suggest $a/M \sim 0.9$, though with significant uncertainty. For an interferometer with $r_1 = 3\,r_s$ and $r_2 = 100\,r_s$, the phase would be $|\Delta\theta| \approx 3.7 \times 10^{24}$ rad.

\textit{M87*}: The BH imaged by the Event Horizon Telescope has mass $M = (6.5 \pm 0.7) \times 10^9\,M_\odot$ and $r_s = 1.92 \times 10^{10}$ km $\approx 128$ AU. Spin estimates from jet modeling suggest $a/M \gtrsim 0.9$. The phase for the same interferometer configuration would be $|\Delta\theta| \approx 5.5 \times 10^{27}$ rad, roughly 1500 times larger than for Sgr A* due to the mass ratio.

\textit{Stellar-mass BH}: A typical X-ray binary BH with $M = 10\,M_\odot$ has $r_s = 29.5$ km. Despite the much smaller gravitomagnetic flux, the proportionally smaller length scales might in principle allow closer approach. The phase for the reference configuration is $|\Delta\theta| \approx 8.5 \times 10^{18}$ rad.

Table~\ref{tab:astro_BH} summarizes the predicted AB phases for these astrophysical systems. The enormous phase values reflect the macroscopic nature of the gravitomagnetic flux near these objects. For Sgr~A*, the Schwarzschild radius of $r_s \approx 1.27 \times 10^7$ km corresponds to roughly $0.085$ AU, while for M87* the horizon scale of $r_s \approx 1.92 \times 10^{10}$ km spans approximately $128$ AU---larger than the orbit of Pluto.

The spin parameters used in Table~\ref{tab:astro_BH} ($a/M = 0.9$) are consistent with current observational constraints. For Sgr~A*, spin estimates from near-infrared flare modeling and stellar orbit analysis suggest $a/M \gtrsim 0.5$, with some analyses indicating $a/M \sim 0.9$ \cite{Genzel:2010,GravityCollaboration:2020}. For M87*, constraints from jet power modeling and EHT observations yield $a/M \gtrsim 0.5$, potentially approaching $a/M \sim 0.9$ \cite{EHT:2019b,Nemmen:2019}. Stellar-mass BHs in X-ray binaries exhibit a range of spins, with several systems showing $a/M > 0.9$ based on continuum fitting and reflection spectroscopy \cite{McClintock:2014,Reynolds:2021}.

The phase values in Table~\ref{tab:astro_BH} scale linearly with both mass and spin, so uncertainties in these parameters propagate directly. For instance, if the true spin of Sgr~A* is $a/M = 0.5$ rather than $0.9$, the predicted phase would be reduced by a factor of $0.5/0.9 \approx 0.56$. Nevertheless, even with conservative parameter choices, the phases remain extraordinarily large compared to any terrestrial quantum interference experiment.

\begin{table}[htbp]
\centering
\caption{Gravitomagnetic AB phase for astrophysical BHs with $a/M = 0.9$, $r_1 = 3\,r_s$, and $r_2 = 100\,r_s$ \cite{Genzel:2010,EHT:2019,McClintock:2014}. Mass values are taken from current best estimates with typical uncertainties of $\sim 10\%$.}
\label{tab:astro_BH}
\begin{tabular}{c|c|c|c|c}
\hline\hline
\rowcolor{orange!50}
Object & $M$ ($M_\odot$) & $r_s$ & Distance & $|\Delta\theta|$ (rad) \\
\hline
Stellar BH (typical) & $10$ & $29.5$ km & $\sim 3$ kly & $8.5 \times 10^{18}$ \\
Sgr A* & $4.3 \times 10^6$ & $1.27 \times 10^7$ km & $26.7$ kly & $3.7 \times 10^{24}$ \\
M87* & $6.5 \times 10^9$ & $1.92 \times 10^{10}$ km & $55$ Mly & $5.5 \times 10^{27}$ \\
\hline\hline
\end{tabular}
\end{table}

\subsection{Phase Sensitivity to BH Spin and Geometric Phase Interpretation}

The linear dependence of the AB phase on spin suggests using interferometric measurements to constrain or measure $a/M$. The fractional uncertainty in the spin determination is related to the fractional phase measurement precision by
\begin{equation}\label{eq:spin_precision}
\frac{\delta a}{a} = \frac{\delta(\Delta\theta)}{\Delta\theta}.
\end{equation}
If the phase can be measured to precision $\delta(\Delta\theta) \sim 1$ mrad (comparable to state-of-the-art SQUIDs), and the total phase is $|\Delta\theta| \sim 10^{22}$ rad for a $10^6\,M_\odot$ BH, then the fractional spin precision would be
\begin{equation}
\frac{\delta a}{a} \sim 10^{-25},
\end{equation}
far exceeding any other known measurement technique.

Of course, this extraordinary precision is purely formal since it assumes the enormous accumulated phase can be measured without loss of coherence or fringe ambiguity. In practice, the measurable quantity would be the phase modulo $2\pi$, and distinguishing among the $\sim 10^{22}$ possible fringe orders would require independent knowledge of the BH parameters to comparable precision. Nevertheless, the gravitomagnetic AB effect provides in principle the most sensitive probe of BH angular momentum at the quantum level.

The gravitomagnetic AB phase admits an elegant interpretation in terms of geometric (Berry) phases \cite{Anandan:1977}. When a quantum system is transported around a closed loop in parameter space, it acquires a phase that depends only on the geometry of the path, not on the rate of transport. The gravitomagnetic AB effect is an example of such a geometric phase, where the ``parameter space'' is the curved spacetime itself.

The phase $\Delta\theta = (m^*/\hbar)\oint g_{t\phi}\,d\phi$ can be rewritten using Stokes' theorem as
\begin{equation}\label{eq:berry}
\Delta\theta = \frac{m^*}{\hbar}\int_{\mathcal{S}} \mathbf{\Omega} \cdot d\mathbf{S},
\end{equation}
where $\mathbf{\Omega} = \nabla \times (g_{t\phi}\hat{\phi})$ is the gravitomagnetic curvature (analogous to Berry curvature) and $\mathcal{S}$ is any surface bounded by the interferometer loop. This formulation emphasizes the topological nature of the effect: the phase depends on the flux of $\mathbf{\Omega}$ through the loop, not on the detailed path taken.

The connection to geometric phases also clarifies why the gravitomagnetic AB effect is gauge-invariant. Just as the Berry phase is independent of the choice of basis for the quantum state, the gravitomagnetic phase is independent of the choice of coordinates for the spacetime. Both reflect the underlying geometry of the relevant space (Hilbert space or spacetime).

}

\section{Observational Prospects and Experimental Challenges} \label{sec5}

The theoretical predictions developed in the preceding sections establish that the gravitomagnetic AB phase for Cooper pairs near a Kerr BH can reach extraordinarily large values, exceeding $10^{22}$ radians for supermassive BHs. However, translating these predictions into observable phenomena faces formidable experimental challenges. In this section, we assess the feasibility of detection by comparing with existing gravitational experiments, analyzing the hostile environment near BHs, examining laboratory analog systems, and identifying the key technological advances required for future observations.

\subsection{Comparison with Existing Gravitomagnetic Measurements}

Frame-dragging effects have been successfully measured in the weak-field regime through precision experiments in Earth orbit. These measurements provide important benchmarks for gravitomagnetic physics and illustrate the sensitivity requirements for detecting related effects.

\textit{Gravity Probe B}: The GP-B satellite mission measured the Lense-Thirring precession of gyroscopes in polar orbit at $642$ km altitude \cite{Everitt:2011}. The frame-dragging precession rate predicted by general relativity is
\begin{equation}\label{eq:LT_precession}
\Omega_{\text{LT}} = \frac{GJ_\oplus}{c^2 r^3} \approx 39~\text{mas/yr},
\end{equation}
where $J_\oplus = 7.07 \times 10^{33}$ kg$\cdot$m$^2$/s is Earth's angular momentum and $r = 7.01 \times 10^6$ m is the orbital radius \cite{Everitt:2011}. The measured value of $37.2 \pm 7.2$ mas/yr confirmed the prediction at the $19\%$ level, limited primarily by unexpected torques from electrostatic patches on the gyroscope surfaces.

\textit{LAGEOS satellites}: Laser ranging to the LAGEOS I and II satellites has measured the Lense-Thirring precession of their orbital nodes \cite{Ciufolini:2004,Ciufolini:2019}. At a semi-major axis of $a \approx 12{,}270$ km, the predicted node precession is approximately $31$ mas/yr. Measurements have confirmed this prediction at the $\sim 10\%$ level, with systematic errors dominated by uncertainties in Earth's gravitational multipole moments.

\textit{LARES satellites}: The more recent LARES and LARES-2 missions, with lower area-to-mass ratios to reduce non-gravitational perturbations, have achieved improved precision in measuring frame-dragging effects \cite{Ciufolini:2023}. These measurements approach the $1\%$ level of accuracy.

The key observation is that all successful gravitomagnetic measurements to date exploit cumulative effects over extended observation periods (years) and rely on classical test masses (gyroscopes, satellites). The gravitomagnetic AB effect for Cooper pairs offers a complementary approach using quantum coherence, but requires access to much stronger gravitational fields than available in the solar system.

\subsection{Tidal Forces and Cooper Pair Coherence}

A fundamental challenge for any superconducting device near a BH is maintaining Cooper pair coherence against tidal disruption. The tidal acceleration across a coherence length $\xi$ at radius $r$ from a BH of mass $M$ is
\begin{equation}\label{eq:tidal_accel}
a_{\text{tidal}} \sim \frac{GM\xi}{r^3} = \frac{\xi c^2}{2r_s} \left(\frac{r_s}{r}\right)^3.
\end{equation}
This must be compared with the binding acceleration scale set by the superconducting condensation energy:
\begin{equation}\label{eq:binding_accel}
a_{\text{bind}} \sim \frac{k_B T_c}{m^* \xi},
\end{equation}
where $T_c$ is the critical temperature and $k_B$ is Boltzmann's constant \cite{Tinkham:2004}.

Table~\ref{tab:tidal_comparison} compares these acceleration scales for representative superconductors near Sgr~A*. The tidal forces become comparable to binding forces only at distances of order $r \sim 10\,r_s$ for conventional superconductors with micron-scale coherence lengths. High-$T_c$ superconductors with nanometer-scale coherence lengths are more resistant to tidal disruption but face other challenges discussed below.

\begin{table}[htbp]
\centering
\caption{Tidal versus binding accelerations for superconductors near Sgr~A* ($M = 4.3 \times 10^6\,M_\odot$, $r_s = 1.27 \times 10^{10}$ m) \cite{Tinkham:2004,Poole:2007,Genzel:2010}. Coherence is maintained when $a_{\text{tidal}} \ll a_{\text{bind}}$.}
\label{tab:tidal_comparison}
\begin{tabular}{c|c|c|c|c}
\hline\hline
\rowcolor{orange!50}
Material & $\xi_0$ & $a_{\text{bind}}$ (m/s$^2$) & $a_{\text{tidal}}$ at $10\,r_s$ & Ratio \\
\hline
Al & $1.6~\mu$m & $5.7 \times 10^{12}$ & $4.5 \times 10^{-13}$ & $7.8 \times 10^{-26}$ \\
Nb & $38$ nm & $1.9 \times 10^{15}$ & $1.1 \times 10^{-14}$ & $5.8 \times 10^{-30}$ \\
YBCO & $1.5$ nm & $4.7 \times 10^{17}$ & $4.3 \times 10^{-16}$ & $9.1 \times 10^{-34}$ \\
\hline\hline
\end{tabular}
\end{table}

The extremely small ratios in Table~\ref{tab:tidal_comparison} indicate that tidal disruption of Cooper pairs is not the limiting factor for superconducting devices near SMBHs at distances $r \gtrsim 10\,r_s$. The situation is more favorable for SMBHs than stellar-mass BHs because, while the total tidal force scales as $M/r^3 \propto 1/r_s^2$, the relevant comparison is made at fixed multiples of $r_s$, where larger BHs produce weaker tidal gradients.

\subsection{Thermal and Radiative Environment}

Maintaining superconductivity requires temperatures below $T_c$, which ranges from $\sim 1$ K for aluminum to $\sim 93$ K for YBCO \cite{Poole:2007}. The thermal environment near a BH presents several heating mechanisms.

\textit{Hawking radiation}: For astrophysical BHs, the Hawking temperature is negligible. For a Kerr BH with mass $M$ and spin parameter $a$, the temperature is \cite{Hawking:1975,Wald:1999vt,Sakalli:2017ewb,Sakalli:2015jaa,Sakalli:2025els}
\begin{equation}\label{eq:Hawking_temp}
T_H^{\text{Kerr}} = \frac{\hbar c^3}{2\pi k_B GM} \cdot \frac{\sqrt{1-(a/M)^2}}{[1+\sqrt{1-(a/M)^2}]^2 + (a/M)^2},
\end{equation}
which reduces to the Schwarzschild result $T_H^{\text{Schw}} = \hbar c^3/(8\pi k_B GM) \approx 6.2 \times 10^{-8}\,\text{K}\,(M_\odot/M)$ in the non-rotating limit ($a = 0$). Crucially, the Kerr temperature is always \textit{lower} than the Schwarzschild temperature for the same mass, vanishing in the extremal limit $a \to M$. For a rapidly spinning BH with $a/M = 0.9$, the ratio $T_H^{\text{Kerr}}/T_H^{\text{Schw}} \approx 0.61$, while for $a/M = 0.99$ it drops to $\approx 0.25$. Thus for Sgr~A* with $M = 4.3 \times 10^6\,M_\odot$ and $a/M \sim 0.9$, we find $T_H \approx 9 \times 10^{-15}$ K, even lower than the Schwarzschild estimate of $1.4 \times 10^{-14}$ K. These temperatures are far below any superconducting $T_c$, so Hawking radiation poses no threat to superconductivity near astrophysical BHs.

\textit{Accretion disk radiation}: Active BHs with accretion disks present a much more hostile thermal environment. For a thin disk around a stellar-mass BH, the effective temperature near the innermost stable circular orbit (ISCO) can reach $T \sim 10^7$ K \cite{Shakura:1973}. For SMBHs, the peak temperature scales as $T \propto M^{-1/4}$ at fixed Eddington ratio, giving $T \sim 10^5$ K for a $10^6\,M_\odot$ BH \cite{Frank:2002}. These temperatures would instantly destroy any superconducting state.

However, quiescent BHs with minimal or no accretion offer more favorable conditions. Sgr~A* is extremely underluminous, with an X-ray luminosity of only $L_X \sim 2 \times 10^{33}$ erg/s in quiescence \cite{Baganoff:2003}, corresponding to a radiative efficiency many orders of magnitude below the Eddington limit. The ambient radiation field at distances $r \gtrsim 100\,r_s$ from a quiescent SMBH may be dominated by the cosmic microwave background at $T_{\text{CMB}} = 2.725$ K \cite{Fixsen:2009}, which is below the $T_c$ of high-temperature superconductors.

\textit{Cosmic ray background}: High-energy cosmic rays can break Cooper pairs through inelastic scattering. The cosmic ray flux at Earth is approximately $1$ particle/cm$^2$/s for energies above $1$ GeV \cite{PDG:2022}. Shielding requirements depend on the detector geometry and acceptable error rates, but are not fundamentally prohibitive for space-based superconducting devices, as demonstrated by missions such as the Alpha Magnetic Spectrometer \cite{AMS:2013}.

\subsection{Comparison with Matter-Wave Interferometry}

The gravitational AB effect has recently been observed using atom interferometry with ultracold rubidium atoms \cite{Overstreet:2022}. This landmark experiment demonstrated that atoms acquire a quantum phase from gravitational potentials in regions where no classical gravitational force acts, directly analogous to the electromagnetic AB effect.

The key parameters differ substantially between atomic and Cooper pair interferometry. The mass coupling for $^{87}$Rb atoms is $m_{\text{Rb}} = 1.44 \times 10^{-25}$ kg, compared to $m^* = 1.82 \times 10^{-30}$ kg for Cooper pairs---a ratio of approximately $8 \times 10^4$. This larger mass enhances the gravitational phase sensitivity, which scales as $m/\hbar$. However, Cooper pairs offer the advantage of macroscopic quantum coherence in SQUIDs, with achievable flux sensitivities of $\sim 10^{-6}\,\Phi_0/\sqrt{\text{Hz}}$ \cite{Clarke:2004}, corresponding to phase sensitivities of $\sim 6 \times 10^{-6}$ rad/$\sqrt{\text{Hz}}$.

The Colella-Overhauser-Werner (COW) experiment \cite{Colella:1975} demonstrated gravitationally induced quantum interference using thermal neutrons ($m_n = 1.67 \times 10^{-27}$ kg) in Earth's gravitational field, achieving phase shifts of order $1$ radian. Subsequent experiments have refined this technique and extended it to rotating reference frames \cite{Werner:1979}, providing direct measurements of the Sagnac effect for matter waves.

Table~\ref{tab:interferometer_comparison} compares the key parameters for different matter-wave interferometry approaches to gravitational phase measurements. The fundamental figure of merit for gravitational sensitivity is the mass-to-Planck-constant ratio $m/\hbar$, which determines the phase accumulated per unit gravitational potential. Heavier particles acquire larger phases for a given potential difference, explaining why atom interferometers achieve better gravitational sensitivity than neutron or electron interferometers despite their shorter de Broglie wavelengths.

The COW experiment \cite{Colella:1975} pioneered gravitational quantum interference by detecting the Earth's gravitational potential using thermal neutrons with wavelengths $\lambda \sim 1.8$ \AA. The phase shift of order $1$ radian arose from the gravitational potential difference across the $\sim 2$ cm height of the interferometer loop, demonstrating that gravity couples to the quantum phase of matter waves exactly as predicted by the equivalence principle \cite{Werner:1979}.

Atom interferometry with ultracold $^{87}$Rb atoms \cite{Overstreet:2022} achieves superior phase sensitivity due to the $\sim 10^2$ larger atomic mass compared to neutrons, combined with advanced laser cooling and manipulation techniques that enable coherent splitting over macroscopic distances ($\sim 25$ cm in the gravitational AB experiment). The demonstrated sensitivity of $\sim 10^{-4}$ rad has enabled the first observation of a gravitational AB effect and opens prospects for improved measurements of Newton's constant $G$ \cite{Roura:2022}.

Cooper pairs, despite their smaller mass ($m^* = 2m_e \approx 1.82 \times 10^{-30}$ kg), benefit from the extraordinary flux sensitivity of SQUIDs, which can resolve phase shifts as small as $\sim 10^{-6}$ rad/$\sqrt{\text{Hz}}$ through their response to the superconducting flux quantum $\Phi_0 = h/(2e)$ \cite{Clarke:2004}. While SQUIDs have not yet detected gravitational phases, their demonstrated electromagnetic AB sensitivity establishes the technical foundation for gravitomagnetic measurements if sufficiently strong gravitational sources become accessible.

\begin{table}[htbp]
\centering
\caption{Comparison of matter-wave interferometry approaches for gravitational phase measurements \cite{Colella:1975,Overstreet:2022,Clarke:2004}. The phase sensitivity scales with the particle mass through the ratio $m/\hbar$, favoring heavier particles for gravitational applications.}
\label{tab:interferometer_comparison}
\begin{tabular}{c|c|c|c|c}
\hline\hline
\rowcolor{orange!50}
System & Mass (kg) & $m/\hbar$ (kg$\cdot$s/J) & Phase sensitivity & Demonstrated effect \\
\hline
Neutrons (COW) & $1.67 \times 10^{-27}$ & $1.58 \times 10^{7}$ & $\sim 0.1$ rad & Gravitational potential \\
Atoms ($^{87}$Rb) & $1.44 \times 10^{-25}$ & $1.37 \times 10^{9}$ & $\sim 10^{-4}$ rad & Gravitational AB \\
Cooper pairs & $1.82 \times 10^{-30}$ & $1.73 \times 10^{4}$ & $\sim 10^{-6}$ rad & EM AB (SQUIDs) \\
\hline\hline
\end{tabular}
\end{table}

\subsection{Laboratory Analog Systems}

While direct observation of the gravitomagnetic AB effect near astrophysical BHs remains beyond current technological capabilities, laboratory analog systems may provide accessible platforms for studying related physics.

\textit{Rotating superconductors}: A rotating superconductor generates a magnetic field through the London moment \cite{London:1950}:
\begin{equation}\label{eq:London_moment}
\mathbf{B}_L = -\frac{2m_e}{e}\boldsymbol{\omega},
\end{equation}
where $\boldsymbol{\omega}$ is the angular velocity. For $\omega = 1000$ rad/s, this gives $B_L \approx 1.1 \times 10^{-8}$ T $= 11$ nT \cite{Tate:1989}. While not directly analogous to gravitomagnetism, the London moment demonstrates how rotation couples to the superconducting order parameter and produces observable macroscopic effects.

\textit{Acoustic black holes}: Unruh's proposal of acoustic (``dumb hole'') analogs \cite{Unruh:1981} has inspired extensive research into analog gravity systems. In a flowing superfluid, the effective metric for phonon propagation can mimic a BH spacetime when the flow velocity exceeds the local sound speed \cite{Visser:1998,Barcelo:2011}. Superfluid helium, with a sound speed of $c_s \approx 238$ m/s, provides a natural medium for such experiments. While these analogs capture aspects of BH physics such as Hawking radiation \cite{Steinhauer:2016}, the gravitomagnetic sector requires rotating flows, which introduce additional experimental challenges.

\textit{Optical analogs}: Slow-light media and metamaterials can create effective curved spacetimes for photons \cite{Leonhardt:2006,Philbin:2008}. While these systems cannot directly probe Cooper pair dynamics, they offer complementary approaches to studying wave propagation in curved effective geometries.

\subsection{Constraints and Limits}

The nearest known stellar-mass BH candidate is V616 Monocerotis (A0620-00), at a distance of approximately $3{,}300$ light-years \cite{Cantrell:2010}. The nearest SMBH is Sgr~A*, at a distance of $\sim 26{,}000$ light-years \cite{GravityCollaboration:2019}. These enormous distances preclude any near-term possibility of deploying instrumentation in the strong-field regime of an astrophysical BH.

Even with optimistic projections for interstellar travel technology, the journey times and energy requirements are prohibitive. A spacecraft traveling at $0.1c$ would require $\sim 33{,}000$ years to reach V616 Mon and $\sim 260{,}000$ years to reach Sgr~A*. The energy required to accelerate a $1000$ kg payload to $0.1c$ is approximately $4.5 \times 10^{20}$ J, equivalent to global energy consumption for several years.

Superconducting Quantum Interference Devices (SQUIDs) represent the most sensitive magnetometers and can detect phase shifts in superconducting loops with extraordinary precision \cite{Clarke:2004}. State-of-the-art DC SQUIDs achieve flux sensitivities of
\begin{equation}\label{eq:SQUID_sensitivity}
\delta\Phi \sim 10^{-6}\,\Phi_0/\sqrt{\text{Hz}},
\end{equation}
where $\Phi_0 = h/(2e) = 2.07 \times 10^{-15}$ Wb is the magnetic flux quantum. This corresponds to a phase sensitivity of $\delta\theta \sim 6 \times 10^{-6}$ rad/$\sqrt{\text{Hz}}$.

For detecting gravitomagnetic effects in the laboratory, the relevant figure of merit is the gravitomagnetic flux that can be generated by realistic rotating masses. A $1$ kg mass rotating at $\omega = 1000$ rad/s at a distance of $1$ m produces a gravitomagnetic potential of order
\begin{equation}
g_{t\phi} \sim \frac{GJ}{c^2 r^2} \sim \frac{G M r^2 \omega}{c^2 r^2} = \frac{GM\omega}{c^2} \sim 7 \times 10^{-25}~\text{m}.
\end{equation}
The resulting phase for a Cooper pair traversing a loop of circumference $2\pi r_{\text{loop}}$ with $r_{\text{loop}} = 10$ cm is
\begin{equation}
\Delta\theta \sim \frac{m^*}{\hbar} \cdot g_{t\phi} \cdot 2\pi r_{\text{loop}} \sim 8 \times 10^{-21}~\text{rad},
\end{equation}
which is $15$ orders of magnitude below the SQUID detection threshold of $\sim 6 \times 10^{-6}$ rad/$\sqrt{\text{Hz}}$. This enormous gap underscores why direct laboratory detection of the gravitomagnetic AB effect for Cooper pairs is not feasible with foreseeable technology. Given the impossibility of direct measurement near astrophysical BHs, alternative approaches merit consideration:

\textit{Gravitational wave signatures}: The inspiral and merger of compact objects produces gravitational waves that encode information about the spacetime geometry, including frame-dragging effects \cite{LIGOScientific:2016}. While not directly probing Cooper pair physics, these observations constrain the Kerr nature of astrophysical BHs and validate the gravitomagnetic potentials that would produce the AB effect.

\textit{Black hole shadow observations}: The Event Horizon Telescope has imaged the shadows of Sgr~A* and M87*, providing direct evidence for the strong-field geometry near SMBHs \cite{EHT:2019,EHT:2022}. The shadow shape and photon ring structure are sensitive to the spin parameter $a/M$, constraining the same gravitomagnetic potential that enters the Cooper pair AB phase.

\textit{Pulsar timing}: Millisecond pulsars in orbit around BHs can serve as precision probes of gravitomagnetic effects through their timing residuals \cite{Wex:2014}. The discovery of a pulsar in close orbit around Sgr~A* would enable tests of frame-dragging in the strong-field regime, complementing the weak-field measurements from GP-B and LAGEOS.

Table~\ref{tab:challenges_summary} summarizes the principal experimental challenges and their severity for observing the gravitomagnetic AB effect with Cooper pairs. The distance barrier is the fundamental obstacle. Even if all other challenges were overcome, deploying a superconducting interferometer within $\sim 100\,r_s$ of an astrophysical BH would require propulsion and materials technologies far beyond current capabilities. The gravitomagnetic AB effect for Cooper pairs in Kerr spacetime thus remains a theoretical prediction whose direct verification awaits revolutionary advances in space exploration, or the discovery of alternative observational signatures that do not require proximity to the source.

\begin{table}[htbp]
\centering
\caption{Summary of experimental challenges for detecting the gravitomagnetic AB effect with Cooper pairs near astrophysical BHs.}
\label{tab:challenges_summary}
\begin{tabular}{c|c|c}
\hline\hline
\rowcolor{orange!50}
Challenge & Severity & Potential mitigation \\
\hline
Distance to BHs & Prohibitive & None foreseeable \\
Accretion disk heating & Severe (active BHs) & Target quiescent systems \\
Tidal disruption & Moderate & Operate at $r > 10\,r_s$ \\
Cosmic ray damage & Manageable & Shielding \\
Hawking radiation & Negligible & Not required \\
Coherence maintenance & Moderate & Low-$T$, short-$\xi$ materials \\
\hline\hline
\end{tabular}
\end{table}


\section{Conclusion} \label{sec6}

We have presented a theoretical analysis of the AB effect for Cooper pairs in the gravitomagnetic field of a Kerr BH. This work extends the well-established electromagnetic AB effect to the gravitational domain, demonstrating that Cooper pairs traversing closed paths around a rotating BH acquire a quantum phase proportional to the enclosed gravitomagnetic flux. The effect is a direct manifestation of frame-dragging at the quantum level and provides a conceptual bridge between superconductivity, quantum mechanics, and general relativity.

The Kerr metric component $g_{t\phi}$ plays the role of a gravitomagnetic vector potential, analogous to the electromagnetic potential $A_\mu$ in the standard AB effect. For a Cooper pair of mass $m^* = 2m_e = 1.82 \times 10^{-30}$ kg completing a circular path in the equatorial plane at radius $r$, the accumulated phase is $\theta = -(4\pi m^* Ma)/(\hbar r)$, where $M$ and $a$ are the BH mass and spin parameter, respectively. The phase difference between two interferometer arms at radii $r_1$ and $r_2$ is given by $\Delta\theta = (4\pi m^* Ma/\hbar)(1/r_2 - 1/r_1)$, which depends only on the enclosed gravitomagnetic flux and not on the local gravitational field experienced by the Cooper pairs---the hallmark of an AB-type topological effect.

Our numerical calculations reveal that the gravitomagnetic AB phase reaches extraordinary magnitudes for astrophysical BHs. For a supermassive BH such as Sgr~A* ($M = 4.3 \times 10^6\,M_\odot$, $a/M = 0.9$), an interferometer with arms at $r_1 = 3\,r_s$ and $r_2 = 100\,r_s$ would accumulate a phase difference of $|\Delta\theta| \approx 3.7 \times 10^{24}$ radians. For M87* ($M = 6.5 \times 10^9\,M_\odot$), the corresponding phase exceeds $10^{27}$ radians. These enormous values reflect the macroscopic scale of the gravitomagnetic potential near astrophysical BHs, with the phase scaling linearly with both mass and spin: $\Delta\theta \propto Ma$.

The structural parallel between electromagnetic and gravitomagnetic AB effects is striking. Both involve quantum phases arising from potentials rather than fields, both are gauge-invariant and topological in nature, and both can be understood through the geometric (Berry) phase formalism. The key difference lies in the coupling constants: electromagnetic phases couple through the charge $e^*$, while gravitomagnetic phases couple through the mass $m^*$. The ratio $m^*/e^* = 5.69 \times 10^{-12}$ kg/C quantifies the relative weakness of gravitational versus electromagnetic coupling at the quantum level.

We have examined the experimental challenges associated with observing this effect near astrophysical BHs. Tidal forces, while potentially destructive to macroscopic structures, pose minimal threat to Cooper pair coherence: at distances $r \gtrsim 10\,r_s$ from Sgr~A*, the tidal acceleration across a coherence length is $\sim 10^{-13}$ m/s$^2$, many orders of magnitude below the binding acceleration scale of $\sim 10^{12}$ m/s$^2$ for aluminum. The thermal environment is similarly benign for quiescent BHs, with Hawking temperatures of $\sim 10^{-14}$ K for SMBHs---far below any superconducting critical temperature. The cosmic microwave background at $T = 2.725$ K would be the dominant thermal input at large distances, still permitting operation of high-$T_c$ superconductors.

The fundamental obstacle to direct observation is the vast distance to astrophysical BHs. The nearest known stellar-mass BH candidate lies approximately $3{,}300$ light-years away, and Sgr~A* is at a distance of $\sim 26{,}700$ light-years. Deploying superconducting instrumentation within the strong-field regime of any BH remains far beyond current or foreseeable technological capabilities. Laboratory detection is equally impractical: a rotating mass in a terrestrial experiment produces gravitomagnetic phases of order $10^{-21}$ radians, some $15$ orders of magnitude below the sensitivity of state-of-the-art SQUIDs.

Nevertheless, the theoretical framework developed here contributes to our understanding of quantum mechanics in curved spacetime and the interplay between gravity and macroscopic quantum coherence. The gravitomagnetic AB effect for Cooper pairs complements recent experimental advances in matter-wave interferometry, including the observation of gravitational AB phases with ultracold atoms \cite{Overstreet:2022} and ongoing efforts to probe quantum superposition in gravitational fields \cite{Belenchia:2018szb}. While the Cooper pair system offers the advantage of macroscopic quantum coherence inherent to superconductors, atom interferometry currently provides greater phase sensitivity due to the larger atomic masses involved.

Our prospective directions are as follows: To begin with, the analysis may be generalized to other BH spacetimes, such as charged (Kerr–Newman) BHs, BHs arising in modified gravity frameworks, and rotating BHs endowed with additional hair or containing exotic matter. The gravitomagnetic AB phase would serve as a probe of the near-horizon geometry in each case. Second, the role of quantum corrections to the Kerr metric, arising from semiclassical gravity or specific quantum gravity proposals, could modify the predicted phases and potentially provide observational signatures of quantum gravitational effects. Third, laboratory analog systems---including rotating superfluids, acoustic BH analogs, and optical metamaterials---may offer accessible platforms for studying related physics, even if they cannot replicate the full gravitomagnetic AB effect.

In summary, we have established that Cooper pairs in the vicinity of a Kerr BH experience a gravitomagnetic AB effect, acquiring quantum phases that encode information about the BH's mass and angular momentum. The effect is topological, gauge-invariant, and represents a genuine quantum signature of frame-dragging. Although direct experimental verification remains beyond current reach, the theoretical predictions underscore the deep connections between quantum coherence and spacetime geometry, and may inform future investigations at the interface of quantum mechanics and general relativity.

}

\section*{Acknowledgments}
We are thankful for academic support provided by EMU, T\"{U}B\.{I}TAK, ANKOS, and SCOAP3, as well as for networking support received through COST Actions CA22113, CA21106, CA23130, and CA23115. E.~S. also acknowledges academic support from Akdeniz University.

\section*{Conflict of Interest}
The authors declare no conflict of interest.

\section*{Data Availability}
No experimental data were generated in this study. All analytical results can be reproduced using the expressions provided.

\bibliography{ref}
\bibliographystyle{unsrt}

\end{document}